\documentclass[]{emulateapj}

\usepackage{natbib}
\usepackage{amsmath}
\usepackage{threeparttable}
\newcommand{\HI}{\ensuremath{\mbox{\rm \ion{H}{1}}}}
\newcommand{\HII}{\ensuremath{\mbox{\rm \ion{H}{2}}}}

\newcommand{\htwo}{\ensuremath{\mbox{H$_2$}}}
\newcommand{\msun}{\ensuremath{M_\odot}}

\newcommand{\zsun}{\ensuremath{Z_\odot}}

\newcommand{\kms}{\mbox{km~s$^{-1}$}}

\newcommand{\av}{\ensuremath{\mbox{$A_{\rm V}$}}}

\newcommand{\mhalo}{\mbox{$M_{\rm halo}$}}
\newcommand{\mdust}{\mbox{$M_{\rm dust}$}}
\newcommand{\mstar}{\mbox{$M_\star$}}
\newcommand{\mgas}{\mbox{$M_{\rm gas}$}}

\newcommand{\tdust}{\mbox{$T_{\rm dust}$}}
\newcommand{\rdust}{\mbox{$R_{\rm dust}$}}
\newcommand{\tcmb}{\mbox{$T_{\rm cmb}$}}
\newcommand{\rgal}{\mbox{$R_{\rm gal}$}}
\newcommand{\snu}{\mbox{$S_\nu$}}

\shorttitle{Dust Across Cosmic Time}
\shortauthors{Imara et al.}

\begin{document}

\title{A Model Connecting Galaxy Masses, Star Formation Rates,\\
and Dust Temperatures Across Cosmic Time}

\author{Nia Imara\altaffilmark{1}, Abraham Loeb\altaffilmark{1}, Benjamin D. Johnson\altaffilmark{1}, 
Charlie Conroy\altaffilmark{1}, and Peter Behroozi\altaffilmark{2}}

\altaffiltext{1}{Harvard-Smithsonian Center for Astrophysics, 60 Garden Street, Cambridge, MA 02138}
\email{nimara@cfa.harvard.edu}
\altaffiltext{2}{Department of Astronomy, University of California at Berkeley, Berkeley, CA, 94720}

\begin{abstract}
We investigate the evolution of dust content in galaxies from redshifts $z=0$ to $z=9.5$.  
Using empirically motivated prescriptions, we model galactic-scale properties---including halo mass, stellar mass, star formation rate, gas mass, and metallicity---to make predictions for the galactic evolution of dust mass and dust temperature in main sequence galaxies.  
Our simple analytic model, which predicts that galaxies in the early Universe had greater quantities of dust than their low-redshift counterparts, does a good job at reproducing observed trends between galaxy dust and stellar mass out to $z\approx 6$.   
We find that for fixed galaxy stellar mass, the dust temperature increases from $z=0$ to $z=6$.  
Our model forecasts a population of low-mass, high-redshift galaxies with interstellar dust as hot as, or hotter than, their more massive counterparts; but this prediction needs to be constrained by observations.  
Finally, we make predictions for observing 1.1-mm flux density arising from interstellar dust emission with the Atacama Large Millimeter Array. 
\end{abstract}



\section{Introduction}
Interstellar dust has a number of important implications for the formation and evolution of galaxies.   
Since high-mass stars produce metals, the building blocks of dust grains in the interstellar medium (ISM), dust abundance is an indicator of the level of star formation activity.  As a byproduct of stellar nucleosynthesis, metals are expelled into the ISM via supernovae and stellar winds, and about 30--50\% of the metals \citep{Draine_2007} condense into dust grains.  Thus, dust traces the metal abundance of galaxies \citep{Lisenfeld_1998, Dwek_1998}.
In addition to being a product of previous star formation, dust also influences the formation of new stars, since it catalyzes the formation of molecular hydrogen (e.g., Gould \& Salpeter 1963), thus enabling the formation of molecular clouds, where stars form.  Moreover, dust contributes to gas cooling \citep[e.g.,][]{Ostriker_1973, Peeples_2014, Peek_2015}, and by stimulating cloud fragmentation, dust may affect the form of the initial mass function \citep{Omukai_2005}.

Besides influencing interstellar chemistry and galaxy physics, dust affects the detectability and observed properties of galaxies.  
Dust grains absorb ultraviolet (UV) light and re-emit the radiation at infrared (IR) wavelengths \citep{Spitzer_1978, Draine_1984, Mathis_1990, Tielens_2005}.  
Since light emitted from galaxies is attenuated by dust, with shorter wavelengths suffering the most attenuation, corrections to the observed spectra are needed to faithfully determine galaxy properties, including the stellar mass and luminosity function.  Particularly at high redshifts, where many surveys are executed in the UV rest frame, the measured properties of galaxies critically depend on dust extinction.  
Because dust strongly extincts optical and UV light, it has especially important consequences for galaxies in the early Universe.  First, it affects the escape fraction of UV photons capable of reionizing the Universe at $z\gtrsim 6$.  Second, conversely, interstellar dust may provide protection from the UV background of the intergalactic medium (IGM) which, by photoionization heating, could have evaporated dwarf galaxies during the epoch of reionization \citep{Barkana_1999}.  Furthermore, dust obscures as much as half of the light in star-forming galaxies \citep{Lagache_2005}, and at high redshifts ($z\gtrsim 3$), our understanding of dust-obscured, star formation activity in typical star-forming galaxies is very incomplete \citep[e.g.,][]{Pope_2017}.

\begin{figure*}[ht]
\centering
\includegraphics[width=5in]{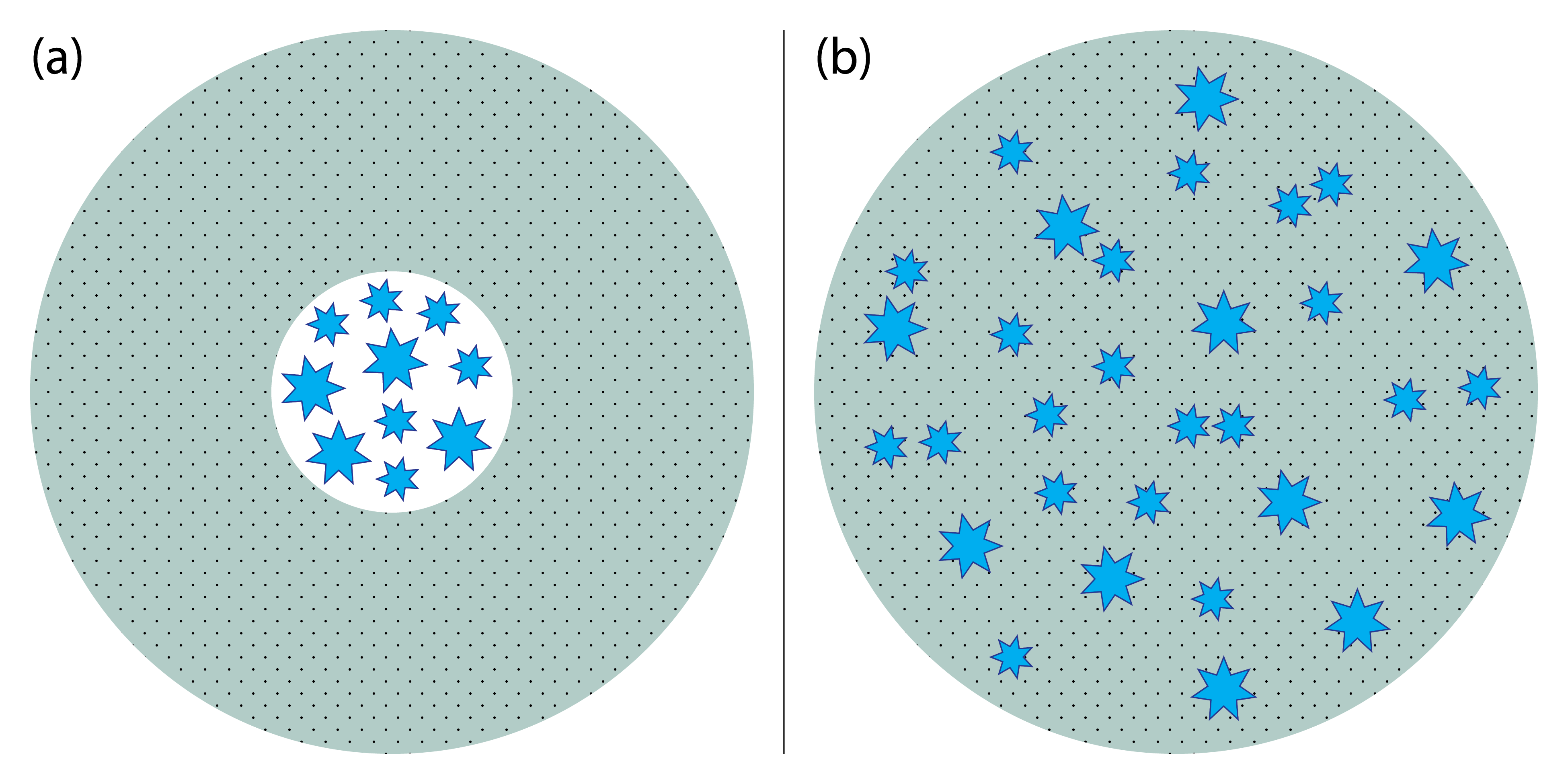}
\caption{Illustration of the two galaxy geometries we consider in this paper.  In panel (a), the stars are located at the galactic center behind a spherically symmetric foreground screen of gas and dust.  Panel (b) represents a homogeneous mixture of stars, dust, and gas.  \label{fig:geometry}}
\end{figure*}
  
The dust content of galaxies in the local and high-redshift Universe has been the focus of a number of observational  studies aiming to understand the physics in the ISM that regulate star formation and constrain galaxy formation models.  The launch of the \emph{Herschel} Space Observatory \citep{Pilbratt_2010} has made possible observational constraints including the relationship between dust mass and stellar mass \citep{Corbelli_2012, Santini_2014}, dust mass and gas fraction \citep{Cortese_2012}, and the evolution of dust temperature \citep{Magdis_2012, Hwang_2010, Magnelli_2014}.  And in recent years, continuum observations with the Atacama Large Millimeter Array (ALMA) have opened a new window on dust formation and evolution in the early Universe.  A number of ALMA programs have detected dust in normal, UV-selected galaxies ($L_{\rm IR} < 10^{12} L_\odot$) from $z=4$--8.4 \citep[e.g.,][]{Capak_2015, Watson_2015, Willott_2015,  Laporte_2017}.  \citet{Dunlop_2017} presented results on the first, deep ALMA image at 1.3-mm of the Hubble Ultra Deep Field.  Particularly exciting are the detections of large amounts of dust in galaxies during the epoch of reionization \citep[e.g.,][]{Watson_2015, Laporte_2017}.  Such observations raise interesting questions about dust production and the rate of supernovae in the early Universe, as significant star formation began at $z\sim 10$--20 \citep{Robertson_2015, Mesinger_2016, Planck_2016}, since the Universe was no more than a few hundred million years old at these redshifts.  The \emph{James Webb Space Telescope} (JWST) is also expected to transform our understanding of dust in the early Universe.

Given this context, the goal of this paper is to investigate the cosmic evolution of galaxy dust mass (\mdust) and dust temperature (\tdust), in ``normal'' star-forming galaxies, with a simple theoretical model.  Such a study is important because measurements of \mdust~and \tdust~shed light on the physical conditions of star-forming environments and are key ingredients in cosmological models of galaxy formation.  A number of groups have used hydrodynamical simulations, analytical models, or semi-analytical models (SAMs), including self-consistent tracking of dust, to make predictions for the evolution of galactic dust \citep[e.g.,][]{Dwek_2007, Dwek_2011, Bekki_2015, McKinnon_2016, Mancini_2016, Popping_2017}.  Traditional simulations and SAMs typically employ recipes for interstellar chemistry and dust production, and they include a great deal of galaxy physics that affect the ISM.  
\citet{Dwek_2007} developed analytical models describing the evolution of high-redshift $(z\gtrsim 6)$ dust, assuming that the evolution of dust depends solely on its production and destruction by core-collapse supernovae.  \citet{Dwek_2011} extended this work by examining the relative roles of supernovae (SNe) and asymptotic branch (AGB) stars in the production of dust by $z\approx 6$.  Both studies were designed to account for the observed dust content in a hyperluminous quasar at $z=6.4$, when the Universe was $\sim 900$ Myr old.

A key advantage of our model that distinguishes it from previous simulations and SAMs is the relative simplicity of its ingredients and, consequently, the comparative facility of physically interpreting the results.  Rather than modeling the micro-physics of galaxies to make predictions about their dust content, we model their large-scale, global properties, including stellar mass, star formation rate (SFR), gas mass, and optical depth due to dust, to determine \mdust~and \tdust.  Our model also has the advantage of producing results useful for observational efforts.  In particular, we make predictions for the evolution of \tdust, a critical quantity for observers interested in estimating the dust mass, especially in high-redshift galaxies.  The total dust mass in a galaxy can only be reliably measured using multi-wavelength observations  of dust emission from IR to sub-millimeter wavelengths, and then modeling the spectral energy distribution (SED).  In the absence of multi-wavelength observations, (for instance, with the recent ALMA observations of high-redshift galaxies), galaxy dust mass is typically estimated by assuming the SED takes the form of a single temperature modified blackbody \citep[e.g.,][]{Watson_2015, Laporte_2017}.  Thus, uncertainties in the total galaxy dust mass and dust-to-gas ratio are usually dominated by the unknown dust temperature.  

In this paper, we use empirically-motivated prescriptions for the relationships between galaxy dark matter halo mass, stellar mass, size, SFR, optical depth due to dust, gas mass, and metallicity to make predictions for the cosmic evolution of \mdust~and \tdust~in normal, main sequence galaxies.

We begin by describing our model for determining the optical depth, mass, and temperature of dust in galaxies in Section \ref{sec:model}.  In Section \ref{sec:results}, we present the results of our model, test them against observations, and make predictions for future infrared observations of high-redshift galaxies.  We discuss caveats and limitations of our model in Section \ref{sec:caveats} and summarize our results in Section \ref{sec:summary}.  
Throughout, we assume a flat, $\Lambda$CDM cosmology with the following parameters: $(\Omega_m, \Omega_\Lambda, \sigma_8)=(0.27, 0.73,0.82)$ and $h=0.7$, where $h$ is the Hubble constant in units of $100$ \kms.

\section{The Model}\label{sec:model}
Our model assumes a simplified, spherical morphology for all main sequence galaxies.  
We consider two geometries for the distribution of stars with respect to the ISM, illustrated in Figure \ref{fig:geometry}.  
In the ``point source'' geometry, the stars are located in the galactic center and are surrounded by a foreground screen of dust and gas.  
The second geometry consists of a homogeneous mixture of stars, dust, and gas. 
In reality, the distribution of dust, gas, and stars is anisotropic and clumpy, with large spiral galaxies having dust, star-forming gas, and high-mass stars concentrated in molecular clouds in the disk.  But a strength of our model is its flexibility, in that it incorporates a variety of galaxies with different morphologies, and  it folds in our poor knowledge of the exact inclination of galaxies, especially at high redshifts.

\begin{table}[t]
\caption{Parameters used in this paper.}\label{table1}
\begin{threeparttable}
\begin{tabular}{lcc}
\tableline\tableline
Symbol   & Definition &      Equation(s)         \\  
\tableline
\tdust        & Dust temperature &  \ref{eq:tdust} \\
$L_\nu$       & Specific luminosity of stars &  \ref{eq:tdust} \\
$f_\star$     & Covering fraction of interstellar dust  &  \ref{eq:tdust} \\
$f_{\rm geom}$ & Geometric factor & \ref{eq:geom}  \\
$\tau_\nu$    & Optical depth due to dust & \ref{eq:tau0}, \ref{eq:tau1}  \\
$r$           & Galactic radius & \ref{eq:rgal}   \\

$r_{\rm vir}$  & Virial radius  &  \ref{eq:rvir}  \\
$\kappa_\nu$  & Dust opacity   &  \ref{eq:kappa}  \\
\mdust        & Dust mass &  \ref{eq:mdust} \\

\mgas         & Gas mass &  \ref{eq:mgas} \\
$\mathcal{Z}$ & Metallicity &  \ref{eq:mz} \\

DGR           & Dust-to-gas ratio & \ref{eq:dgr}  \\

\mstar        & Stellar mass  &   \\
\mhalo        & Halo mass &   \\
\tableline
\end{tabular}
\end{threeparttable}
\end{table}

To describe the redshift evolution of dust temperature in an individual galaxy, $\tdust(z)\equiv\tdust$, we assume the dust is in thermal equilibrium with the total radiation field of the galaxy.  That is, dust grains emit and absorb energy at the same rate,
\begin{equation}
\frac{dE_{\rm emit}}{dt} = \frac{dE_{\rm absorb}}{dt}.
\end{equation}
We assume that the stellar radiation field and the cosmic microwave background (CMB) contribute to dust heating \citep[e.g.,][]{Rowan-Robinson_1979, DaCunha_2013} and that dust grains cool via blackbody radiation.  Thus, the power per unit mass emitted by dust is equal to the total power absorbed per unit mass of dust, according to 
\begin{equation}\label{eq:tdust}
\begin{split}
\int_0^\infty \frac{8\pi h}{c^2}\frac{\nu^3}{{\rm exp}(h\nu/k\tdust)-1}\kappa_\nu d\nu = \\
\int_0^\infty \frac{ L_\nu}{r^2} f_{\rm geom} f_\star \kappa_\nu d\nu \\
 + \int_0^\infty \frac{8\pi h}{c^2}\frac{\nu^3}{{\rm exp}(h\nu/k\tcmb)-1}\kappa_\nu d\nu,
\end{split}
\end{equation}
where $\tcmb=2.725(1+z)$ is the CMB temperature at a given redshift $z$,  $L_\nu$ is the specific luminosity of all the stars in a galaxy, $r$~is the galactic radius, and $\kappa_\nu$ is the dust opacity at frequency $\nu$.  
To account for the porosity of the ISM and the fact that some stellar radiation will escape a galaxy without being absorbed by dust grains, we include the factor $f_\star$, a number between 0 and unity that parameterizes the fraction of a galaxy's surface area covered by dust.  
The factor $f_{\rm geom}$ accounts for the geometry of stars and dust, as
\begin{equation}\label{eq:geom}
 f_{\rm geom} = 
  \begin{cases} 
    e^{-\tau_\nu}               &   \text{Case 1: point source} \\
    (1-e^{-\tau_\nu})/\tau_\nu  &   \text{Case 2: homogeneous}, 
  \end{cases}
\end{equation}
where $\tau_\nu$ is the optical depth due to dust.  The first case corresponds to the case in which the stars at the galactic center act, in effect, as a single point source of radiation behind a foreground screen of dust.  The second case corresponds to the solution of the radiative transfer equation for a homogeneous mixture of stars and dust \citep{Mathis_1972, Natta_1984}.

In order to solve equation (\ref{eq:tdust}) for \tdust, we model a number of parameters for each galaxy at a given redshift, including \rgal, $\kappa_\nu$, $\tau_\nu$, and $L_\nu$.  These parameters and others used in this paper are summarized in Table \ref{table1}.

\subsection{Dust Optical Depth and Mass}\label{sec:tau}

For each individual galaxy, we assume a spherically symmetric system in which the dust mass density, $\rho_{\rm dust}$, 
has a power-law distribution, $\gamma$.  
The frequency-dependent optical depth due to dust is defined

\begin{equation}\label{eq:tau0}
\begin{split}
\tau_\nu &= \int_0^{\rdust} \rho_{\rm dust}(r) \kappa_\nu dr \\
         &= \int_0^{\rdust} \rho_0 r^{-\gamma} \kappa_\nu dr,
\end{split}
\end{equation}
where $\rho_0$ is the central mass density of a given galaxy, and \rdust~is the radial extent of dust. The total dust mass in a galaxy also depends on $\rho_0$ as
\begin{equation}\label{eq:dust_mass}
\begin{split}
M_{\rm dust} &= \int_0^{\rdust} \rho_{0} r^{-\gamma} 4\pi r^2 dr \\
            &= 4\pi\rho_0 \frac{\rdust^{3-\gamma}}{3-\gamma},
\end{split}
\end{equation}
By equating equations (\ref{eq:tau0}) and (\ref{eq:dust_mass}), we may express $\tau_{\rm dust,\nu}$ in terms of dust mass:
\begin{equation}\label{eq:tau1}
\tau_{\nu} = \frac{M_{\rm dust}(M_\star, z)}{4\pi \rdust^2}\frac{3-\gamma}{1-\gamma}\kappa_\nu,
\end{equation}
where we have drawn attention to the dependency of \mdust~on the stellar mass of the galaxy, \mstar, and on the redshift, $z$.  We let $\gamma =0$; we justify this choice in \S\ref{sec:caveats}.  To evaluate equation (\ref{eq:tau1}), then, we need to model $r$, $\kappa_\nu$, and \mdust, all of which depend on $z$.

\subsubsection{Galactic radius}\label{sec:radius}
To acquire a rough approximation of galaxy disk sizes, we adopt the basic picture of \citet{Fall_1980} and others \citep[e.g.,][]{Mo_1998, Somerville_1999}, in which the collapsing gas of a forming galaxy acquires the same specific angular momentum as the dark matter halo, and this angular momentum is conserved as the gas cools.  The specific angular momentum is often expressed using the dimensionless spin parameter, $\lambda=J|E|^{1/2}G^{-1}M^{-5/2}$, where $J$ is the angular momentum, $E$ is the total energy of the halo, $G$ is Newton's gravitational constant, and $M$ is the mass \citep[e.g.,][]{Peebles_1969, Mo_1998, Somerville_1999}.  For a halo with a singular isothermal density profile $\rho\propto r^{-2}$, the disk exponential scale radius is given by $r_d = \lambda R_{\rm halo}/\sqrt{2}$, where $R_{\rm halo}$ is the virial radius of the dark matter halo.

\citet{Somerville_2018} explored $\lambda$ using empirical constraints from $z\sim 0.1$--3 galaxies in the GAMA survey \citep{Driver_2011, Liske_2015} and CANDELS survey \citep{Grogin_2011, Koekemoer_2011}.  Using relationships from their halo abundance matching model, they mapped galaxy stellar mass to halo mass and inferred from these results a median value of the spin parameter, $\lambda=0.036$, corresponding to a ratio between galaxy half-mass radius and halo size of $\rgal/R_{\rm halo} = 0.018$.  They found that $\lambda$ is roughly independent of stellar mass and exhibits weak dependence on redshift.  We adopt this value for $\lambda$ and make the simplifying assumption that the radial extent of the dust, \rdust, is equal to $\rgal$,
\begin{equation}\label{eq:rgal}
\rdust=\rgal=0.018 R_{\rm halo}.
\end{equation}
In \S\ref{sec:caveats}, we discuss some of the limitations of assuming a single value for the spin parameter.

\begin{figure}[t]
\epsscale{1}
\plotone{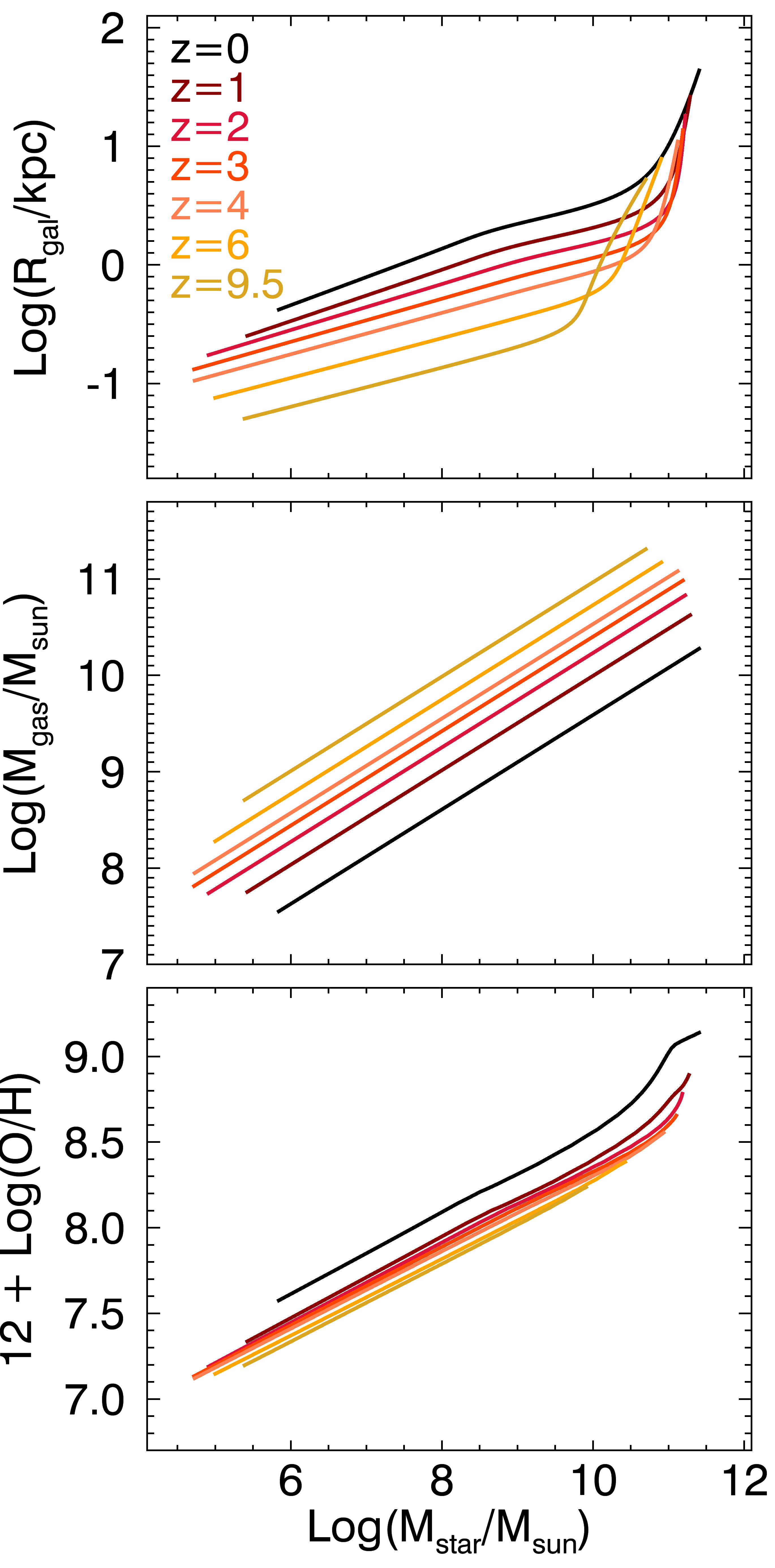}
\caption{Basic properties of the model.  From top to bottom, the galactic radius, gas mass, and metallicity as a function of stellar mass.  \label{fig:model}}
\end{figure}

To determine  $R_{\rm halo}$, we adopt the $R_{\rm halo}$-$M_{\rm halo}$ relation  of \citet{Loeb_2013}:
\begin{equation}\label{eq:rvir}
\begin{split}
R_{\rm halo} &= 0.784 \left[ \frac{\Omega_m}{\Omega_m(z)} \frac{\Delta_c}{18\pi^2} \right]^2 \\
&\times \left( \frac{M_{\rm halo}}{10^8~M_\odot}^{1/3}\right) \left(\frac{10}{1+z}\right) h^{-2/3}~\rm{kpc},
\end{split}
\end{equation}
where 
\begin{equation}
\begin{split}
\Delta_c &= 18\pi^2 + 82d - 39d^2 \\
d &= \Omega_m(z)-1 \\
\Omega_m(z) & = \frac{\Omega_m (1+z)^3}{\Omega_m(1+z)^3 + \Omega_\Lambda}.
\end{split}
\end{equation}
For \mhalo~in equation (\ref{eq:rvir}), we employ the stellar-mass-halo-mass (SMHM) relations of \citet{Behroozi_2013a}, discussed in detail in \S\ref{sec:smhm}.  In Figure \ref{fig:model}, we show how the $\rgal$-$\mstar$ relation evolves with redshift.

\subsubsection{Opacity}\label{sec:opacity}

For most extragalactic environments, especially for distant galaxies, we have poor knowledge of the composition and size distribution of dust grains, encapsulated in $\kappa_\nu$ in equation (\ref{eq:tau1}).  For values of $\nu>10^{12}$ Hz,  we adopt the Galactic extinction laws of \citet{Mathis_1990} and \citet{Li_2001}.  We calculate an interpolated function for $\kappa_\nu$ based on a combination of these two models, since the \citet{Mathis_1990} model extends to lower frequencies, down to $\sim 10^{12}$ Hz, while the \citet{Draine_2001} model extends up to $\sim 10^{18}$ Hz.  For frequencies below $\nu \le 10^{12}$ Hz, we adopt the Beckwith et al. (1990) power-law treatment for opacity, 
\begin{equation}\label{eq:kappa}
\kappa_\nu=0.1 \left( \frac{\nu}{1000~\rm{GHz}} \right)^\beta~\rm{cm}^2 \rm{g}^{-1},
\end{equation}
where $\beta=2$.  The Beckwith et al. model, which originally assumed $\beta=1$, was calibrated to match the emmisivity properties of dust around Galactic protoplanetary disks.  The frequency-dependence of $\kappa_\nu$ is uncertain and depends on the size and composition of dust grains.    Using a different slope or normalization for  $\kappa_\nu$ \citep[e.g.,][]{James_2002, Dunne_2011, daCunha_2008, Clark_2016} would affect the calculation of the optical depth (equation \ref{eq:tau1}) in our model and thus the dust temperature.  We describe how changes in the normalization or slope for $\kappa_\nu$ affect our results in Section \ref{sec:omd}. 

\begin{figure*}
\centering
\includegraphics[width=6in]{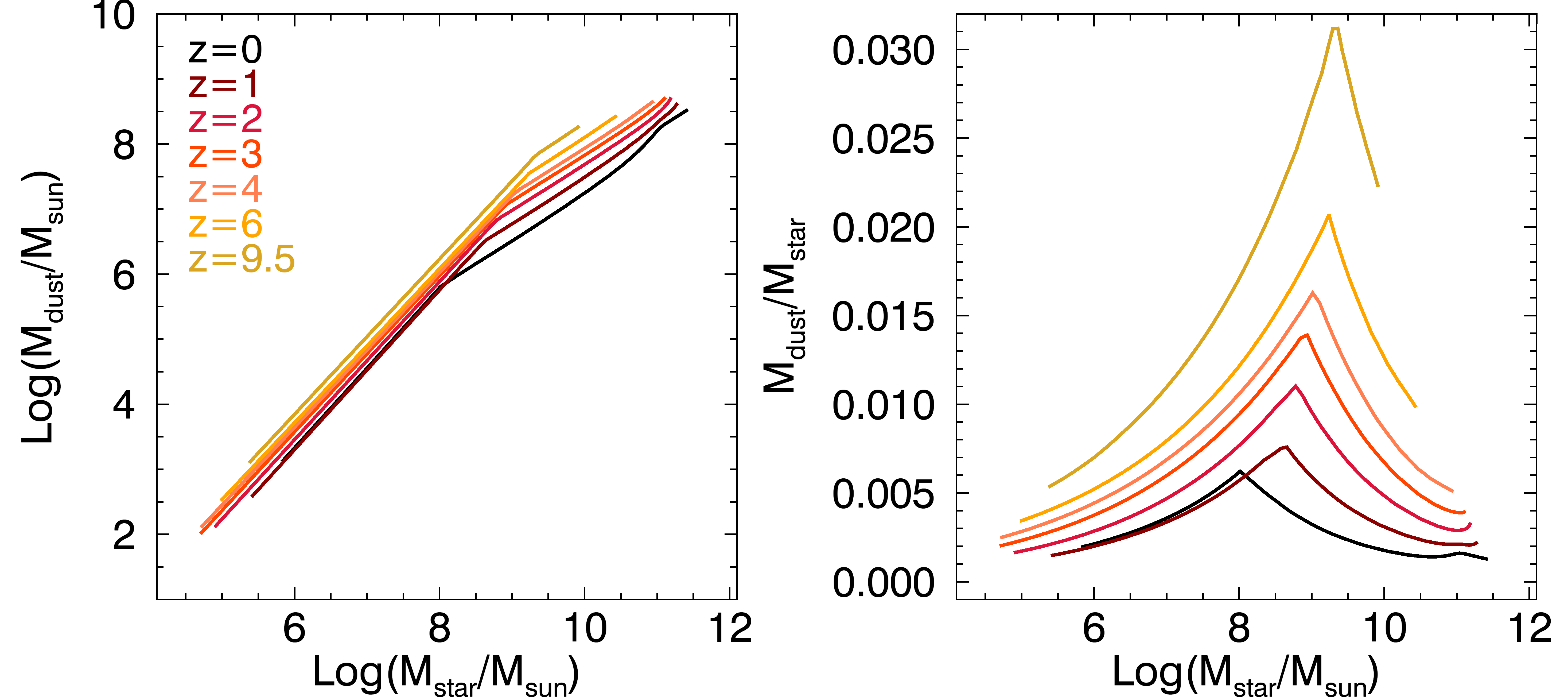}
\caption{Galaxy dust mass versus stellar mass (left panel) and $\mdust/\mstar$ versus \mstar~(right panel), for different redshifts.  \label{fig:mdust}}
\end{figure*}

\subsubsection{Dust Mass}
Next, we determine \mdust, which relates to the gas mass of a galaxy, \mgas, via a dust-to-gas ratio, DGR:
\begin{equation}\label{eq:mdust}
\mdust \equiv \mgas\times\rm{DGR}.
\end{equation}
Several studies have explored the correlation between galactic gas mass and stellar mass or SFR \citep[e.g.,][]{Sargent_2014, Zahid_2014}.  
Yet empirical relations for the total gas mass as a function of stellar mass or SFR rarely extend to redshifts higher than $z\approx 2$ \citep[e.g.,][]{Zahid_2014}, beyond which, measurements of the \HI~gas mass are unreliable, and the gas fraction of galaxies is expected to be increasingly dominated by molecular gas.

After investigating different prescriptions for the gas mass, we decided to follow \citet{Zahid_2014}, who fit a stellar-mass-metallicity (MZ) relation for star-forming galaxies at $z\lesssim 1.6$.  They assume that $\mstar/\mgas \approx (\mstar/M_0)^\gamma$, where $M_0$ is a metallicity-dependent, characteristic mass, above which $\mathcal{Z}$ approaches a saturation limit, and $\gamma$ is a power law index.  By combining this expression with their fit for the MZ relation, \citet{Zahid_2014} determine
\begin{equation}\label{eq:mgas}
\mgas(\mstar,z) = 3.87\times 10^9 (1+z)^{1.35}\left( \frac{\mstar}{10^{10}~\msun} \right)^{0.49}.
\end{equation}
To determine the stellar mass in the above equation, we use the SMHM models of \citet{Behroozi_2013a}, who constrain average galaxy stellar masses and SFRs as a function of halo mass (see \S\ref{sec:smhm}).  We extrapolate the $M_{\rm gas}$-\mstar~relation of Zahid et al. (2014) to redshifts $z> 1.6$, as shown in Figure \ref{fig:model}.  We discuss potential consequences of this choice and investigate an alternative prescription for \mgas~in \S\ref{sec:caveats}. 

Equation (\ref{eq:mdust}) also depends on the dust-to-gas ratio, DGR.  Since dust is composed of heavy elements and traces the metal abundance of galaxies, the problem of determining the DGR can be reduced to one of determining the metallicity, $\mathcal{Z}$.  Several authors have conducted observational investigations of the MZ relation in nearby galaxies \citep[e.g.,][]{Lequeux_1979, Lee_2006, Zahid_2012, Berg_2012, Zahid_2014} and in distant galaxies out to $z\lesssim 3$ \citep{Savaglio_2005, Erb_2006, Maiolino_2008, Yabe_2012, Zahid_2013, Hunt_2016}.  We adopt the prescription of \citet{Hunt_2016}, who compiled observations of $\sim 1000$ galaxies up to $z\sim 3.7$, with metallicities spanning two orders of magnitude, SFRs spanning 6 orders of magnitude, and stellar masses spanning 5 orders of magnitude.  Using a principal component analysis, Hunt et al. (2016) find
\begin{equation}\label{eq:mz}
\mathcal{Z} = -0.14 \log(\rm{SFR})+0.37 \log(\mstar) +4.82,
\end{equation}
where, by convention, $\mathcal{Z}\equiv 12 + \log(\rm{O/H})$ is defined in terms of the gas-phase oxygen abundance (O/H).  We extrapolate equation (\ref{eq:mz}) for galaxies at $z>3.7$, as shown in Figure \ref{fig:model}.

We now relate $\mathcal{Z}$ to the DGR, using an empirical formula determined for local galaxies.  
R\'emy-Ruyer et al. (2014) evaluated the gas-to-dust ratio as a function of metallicity 
for nearby galaxies spanning the range between $1/50~\mathcal{Z}$ and $2~\mathcal{Z}$. 
The authors provide alternative
functional forms for the gas-to-dust ratio versus
metallicity relationship, depending on the CO-to-\htwo~conversion 
factor they employed to estimate the total amount
of molecular mass in a galaxy. We adopt the function
they derive assuming a metallicity-dependent conversion
factor (as opposed to the standard Milky Way CO-to-\htwo~conversion factor).  
In terms of the DGR,
\begin{equation}\label{eq:dgr}
 \log\left(\frac{\mathrm{DGR}}{\mathrm{DGR}_\odot}\right) = 
  \begin{cases} 
   \log \left(\frac{\mathcal{Z}}{\zsun}\right)   &     \text{if } \mathcal{Z} > 0.26\zsun \\
   3.15\log \left(\frac{\mathcal{Z}}{\zsun}\right) + 1.25 & \text{if } \mathcal{Z} \le 0.26\zsun ,
  \end{cases}
\end{equation}
where $\log(\mathrm{DGR}_\odot)=-2.21$ (Zubko et al. 2004).

In the previous subsections, we have modeled $r$, $\kappa_\nu$, \mgas, and the DGR.  
Equation (\ref{eq:rvir}) depends on \mhalo,  equation (\ref{eq:mgas}) on \mstar,  and equation (\ref{eq:mz}) on \mstar~and the SFR  In the next section, we describe the self-consistent models of Behroozi et al. (2013a) that we use to determine relation between \mhalo, \mstar, and the SFR at arbitrary redshifts.

\begin{table*}[t]
\centering
\caption{Parameters derived in this paper for galaxies having $\mhalo=10^{12}$ \msun.}\label{table2}
\begin{threeparttable}
\setlength{\tabcolsep}{12pt}
\begin{tabular}{lccccccc}
\tableline\tableline
                    & $z=0$    & $z=1$    & $z=2$    & $z=3$    & $z=4$    & $z=6$    & $z=9.5$ \\
\tableline
\mstar~($10^{10}$ \msun)  &  2.7     &   2.8    &   2.2    &  2.1     &   2.3    &   1.7    &   0.79  \\
SFR (\msun~yr$^{-1}$) &  0.76    &   15     &  32      &  52      &   82     &   88     &  59     \\
\rgal~(kpc)          &  4.7     &   2.8    &   2.0    &   1.5    &   1.2    &  0.85    &  0.57   \\
$\mathcal{Z}$        &  8.7     &    8.5   &   8.4    &   8.4    &  8.4     &  8.3     &   8.2   \\
DGR                  &  1/160   &   1/241  &  1/292   &  1/315   &  1/326   &  1/367   &  1/463  \\
\av~(mag)            &  0.79    &   1.6    &   2.6    &  4.5     &    7.2   &   11     &  14     \\
\mgas~($10^{10}$ \msun) &  0.6  &   1.6    &   2.5    &   3.7   &   5.1     &   7.0    &  8.3     \\
\mdust~($10^7$ \msun)&  3.9     &   6.7    &   8.5    &   12     &   16     &   19     &  18     \\
\tdust~(K)           &  34      &   51     &   57     &   58     &   60     &   58     &  46     \\
\tableline
\end{tabular}
\textbf{Note.} From top to bottom: stellar mass, SFR, half-mass radius, metallicity, dust-to-gas ratio, visual extinction due to dust, gas mass, dust mass, and dust temperature.
\end{threeparttable}
\end{table*}


\subsection{Stellar-Mass-Halo-Mass Relation}\label{sec:smhm}
\citet{Behroozi_2013a} use empirical forward modeling to constrain the evolution of the stellar mass—halo mass relationship (SMHM; $SM(M_h, z)$).  At fixed redshift, the adopted model for $SM(M_h,z)$ has six parameters, which control the characteristic stellar mass, halo mass, faint-end slope, massive-end cutoff, transition region shape, and scatter of the SMHM relationship.  For each parameter, there are three variables that control its redshift scaling at low ($z=0$), mid ($z=1$--2), and high ($z>3$) redshift, with constant (i.e., no) scaling beyond $z=8.5$ to prevent unphysical early galaxy formation.  Additional nuisance parameters include systematic uncertainties in observed galaxy stellar masses and SFRs.  Any choice of model in this parameter space gives a mapping from simulated dark matter halo catalogs \citet{Behroozi_2013b} to mock galaxy catalogs.  Comparing these mock catalogs with observed galaxy number counts and SFRs from $z=0$ to $z=8$ results in a likelihood for a given model choice, and so these constraints combined with an MCMC algorithm result in a posterior distribution for the allowed SMHM relationships.  Average star formation rates and histories for galaxies are inferred from averaged halo assembly histories (including mergers) combined with the best-fitting model for $SM(M_h,z)$.

\begin{figure}[h]
\epsscale{1.}
\plotone{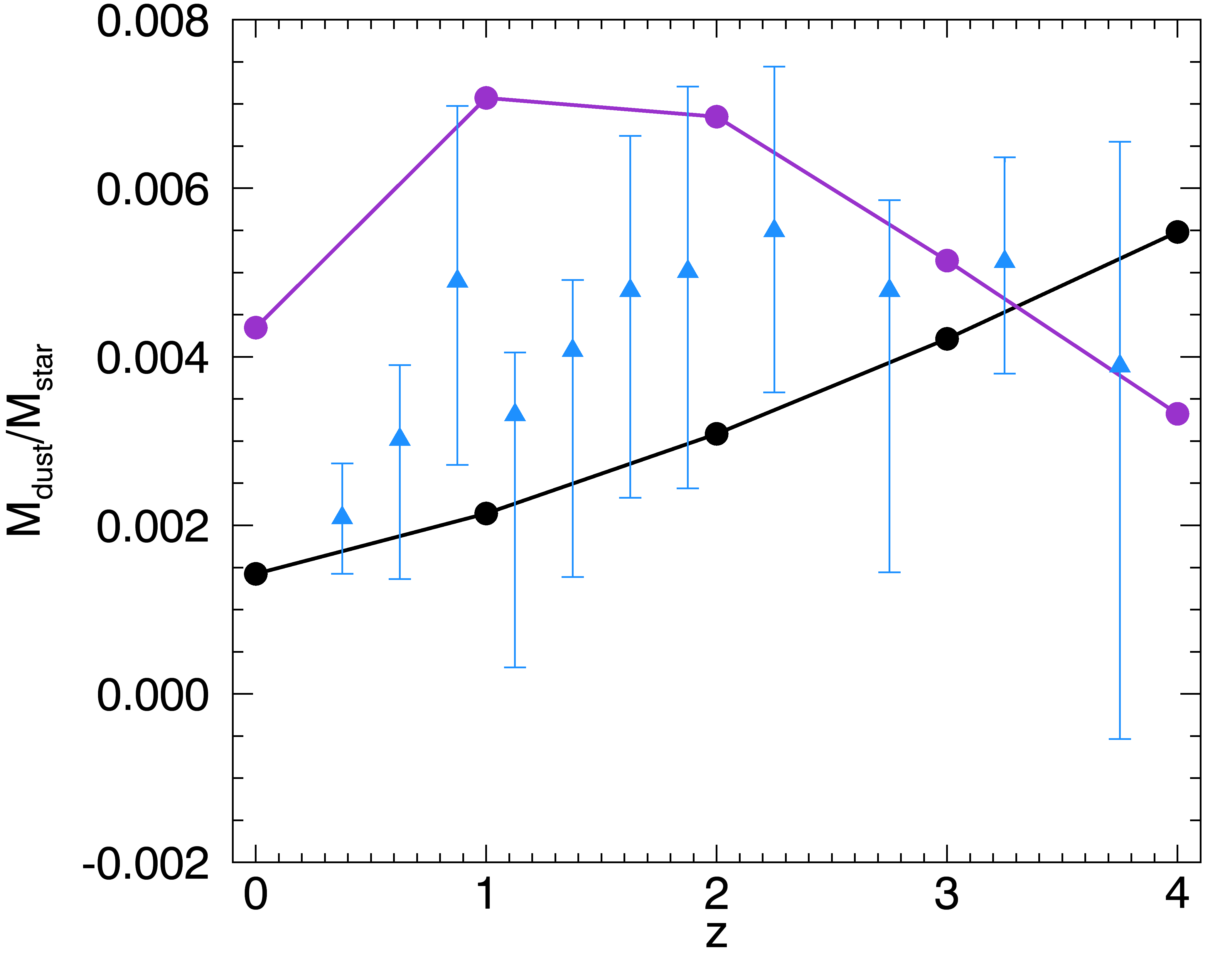}
\caption{Galaxy dust-to-stellar mass ratio as a function of redshift for galaxies with $\mstar = 6\times 10^{10}\msun$.  The black line represents $\mdust/\mstar$ derived in our fiducial model.  The blue triangles are the values measured by \citet{Bethermin_2015} for a sample of main-sequence galaxies. The purple line shows the ratio using the same metallicity-gas mass relation as \citet{Bethermin_2015}, and it includes only star-forming galaxies.  \label{fig:mdustz}}
\end{figure}

\subsection{Stellar Population Synthesis}

To determine the specific luminosity of all the stars in a galaxy, $L_\nu$, we use version 3.0 of the Flexible Stellar Population Synthesis code \citep[FSPS;][]{Conroy_2009, Conroy_2010} to model the spectral energy distributions from 91\AA to 1000\micron.
For each halo and each redshift we supply the star formation history (SFH) to FSPS in tabular form, including the the time-dependent metallicity of newly born stars.
Within FSPS, the SFH is linearly interpolated between the supplied time points, and the appropriate single-stellar population (SSP) weights are calculated.  
The spectra of the SSPs are then summed with these weights applied to produce a model galaxy spectrum corresponding to the last time-point of the supplied SFH.
This model spectrum is also projected onto filter transmission curves to produce broadband rest frame photometry in several standard filter sets.
The total surviving stellar mass (including remnants) and bolometric luminosity are also calculated. 
No dust attenuation or IGM attenuation is applied, and we do not include nebular emission from \HII~regions or dust emission.

The base SSP spectra are generated assuming a fully sampled Salpeter IMF from 0.08 to 120 \msun.   
We use the ``Padova2007'' isochrones \citep{Bertelli_1994, Girardi_2000, Marigo_2008} for stars less than 70 \msun.  
These are combined with Geneva isochrones for higher-mass stars based on the high mass-loss rates evolutionary tracks \citep{Schaller_1992, Meynet_2000} and the post-AGB evolutionary tracks of \citet{Vassiliadis_1994}.
For the stellar spectra we use the BaSeL3.1 theoretical stellar library of \citet{Westera_2002}, augmented with the higher-resolution empirical MILES library \citep{Sanchez-Blazquez_2006} in the optical.
The spectra of OB stars are from Smith et al. (2002), and the spectra of post-AGB stars are from \citet{Rauch_2003}.
The treatment of TP-AGB spectra and isochrones is described in \citet{Villaume_2015}.
All FSPS variables that affect the isochrones and stellar spectra are set at their default values.

\section{Results and Discussion}\label{sec:results}

We now present predictions for the evolution of dust in galaxies from redshifts $z=0$ to $z=9.5$.  
The galaxies have dark matter halo masses ranging from $10^9$ to $10^{15}~\msun$, stellar masses ranging from $4.3\times 10^4$ to $1.6\times 10^{12}~\msun$, and SFRs from 0 to 155 \msun yr$^{-1}$, with average star formation rates taken as a function of halo mass and redshift from Behroozi et al. (2013).  We here consider only average population results, leaving starbursts \citep[e.g.,][]{Riechers_2013, Strandet_2017} and the distribution of dust properties for follow-up studies.  In addition, the \citet{Behroozi_2013a, Behroozi_2013b} analysis does not separate star-forming from quiescent galaxies, constraining only the average SFR of the entire galaxy population as a function of halo mass.

\subsection{Evolution of Dust Mass}

Figure  \ref{fig:mdust} presents galactic dust mass as a function of stellar mass from $z=0$ to $z=9.5$.  We find that for a fixed galactic stellar mass, \mdust~increases with increasing redshift, with higher mass galaxies displaying slightly larger increases in \mdust~than their lower mass counterparts.  In Table \ref{table2}, we summarize the predicted values of \mdust, as well as other parameters derived in this paper, for a   $10^{12}$ \msun~galaxy---to show that our predictions compare favorably with quantities observed in the Milky Way.

\begin{figure*}
\centering
\includegraphics[width=6in]{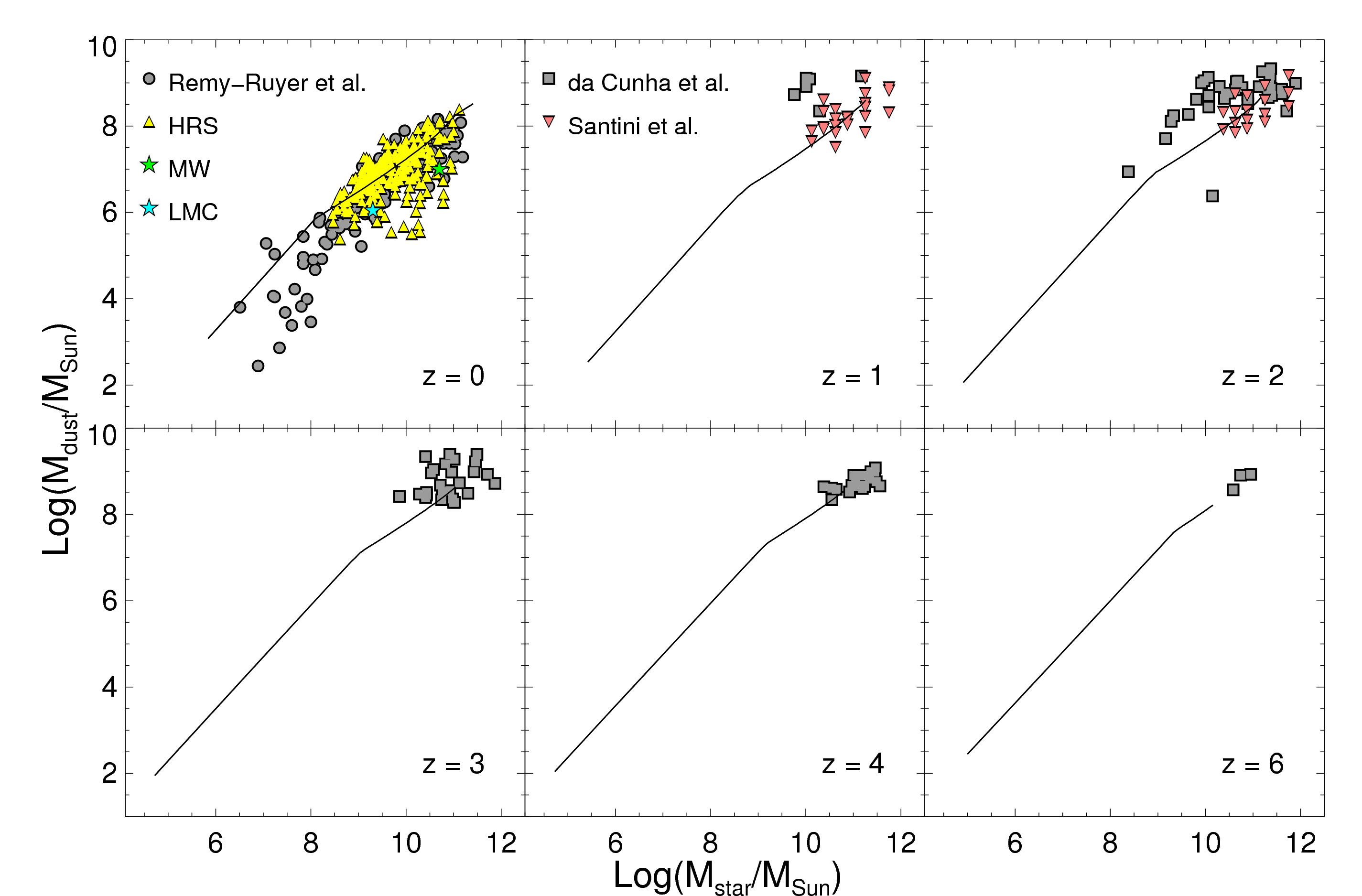}
\caption{Galaxy dust mass versus stellar mass for different redshifts.  The green and turquoise stars represent values for the Milky Way and the Large Magellanic Cloud, respectively.  The triangles are measurements from the Herschel Reference Survey.   \label{fig:mdust_multi}}
\end{figure*}

As the second panel is Figure \ref{fig:mdust} shows, the dust-to-stellar mass ratio  $\mdust/\mstar$ first rises, and then at a characteristic value of \mstar, the ratio decreases with increasing stellar mass.  This turnover results from the dependency of \mdust~on metallicity (via the DGR), which changes its functional form at $\mathcal{Z}=0.26$ (equation \ref{eq:dgr}).  The ratio $\mdust/\mstar$ arises from a competition between the gas metallicity decreasing with redshift and the SFR increasing.  For a fixed stellar mass, $\mdust/\mstar$ steadily increases with redshift.    In Figure \ref{fig:mdustz}, we compare our results with the observations of \citet{Bethermin_2015}, who measure the gas and dust content of massive ($\sim 6\times 10^{10}\msun$) main-sequence galaxies  and find little systematic variation of the dust-to-stellar mass ratio as a function of redshift up to $z=4$.  

Although our results slightly underpredict the observations and suggest a small increase in $\mdust/\mstar$ over the same redshift range, they are compatible with the B\'ethermin et al. observations at the $1\sigma$ level.  There are various ways to explain the lower dust-to-stellar mass ratios: low metallicities and dust-to-gas ratios, or differences in the SFRs and gas masses.  It happens that the gas masses we derive for $\sim 6\times 10^{10}\msun$ galaxies are in good agreement with the gas masses measured by \citet{Bethermin_2015}.  However, if we use a different prescription for the metallicity, this would alter our results for \mdust.  For instance, the fundamental metallicity relation of \citet{Mannucci_2010} yields higher metallicities than the Hunt et al. (2016) prescription we use here.  Moreover, the SFRs derived by B\'ethermin et al. for their galaxy sample are slightly higher than the SFRs we use from the \citet{Behroozi_2013a, Behroozi_2013b} model, since the latter considers \emph{average} SFRs including contributions from both star-forming and quiescent galaxies.  For a fixed stellar mass, a higher SFR results corresponds to a lower metallicity.  In Figure \ref{fig:mdustz} we plot the redshift evolution $\mdust/\mstar$, where $\mdust$ is derived using the metallicity prescription of \citet{Mannucci_2010} and where the SFRs are corrected to include star-forming galaxies only.  This correction is accomplished by dividing the SFRs by $(1-f_q)$, where $f_q$, the fraction of quiescent galaxies at a given redshift, is given by $f_q = [(\mstar/10^{10.2+0.5z}\msun)^{-1.3}+1]^{-1}$ \citep{Behroozi_2013a}.  The resulting curve for the evolution of $\mdust/\mstar$  now slightly overpredicts the  B\'ethermin et al. observations at $z<3$ and does not vary monotonically with redshift.

\begin{figure*}[t]
\centering
\includegraphics[width=6.5in]{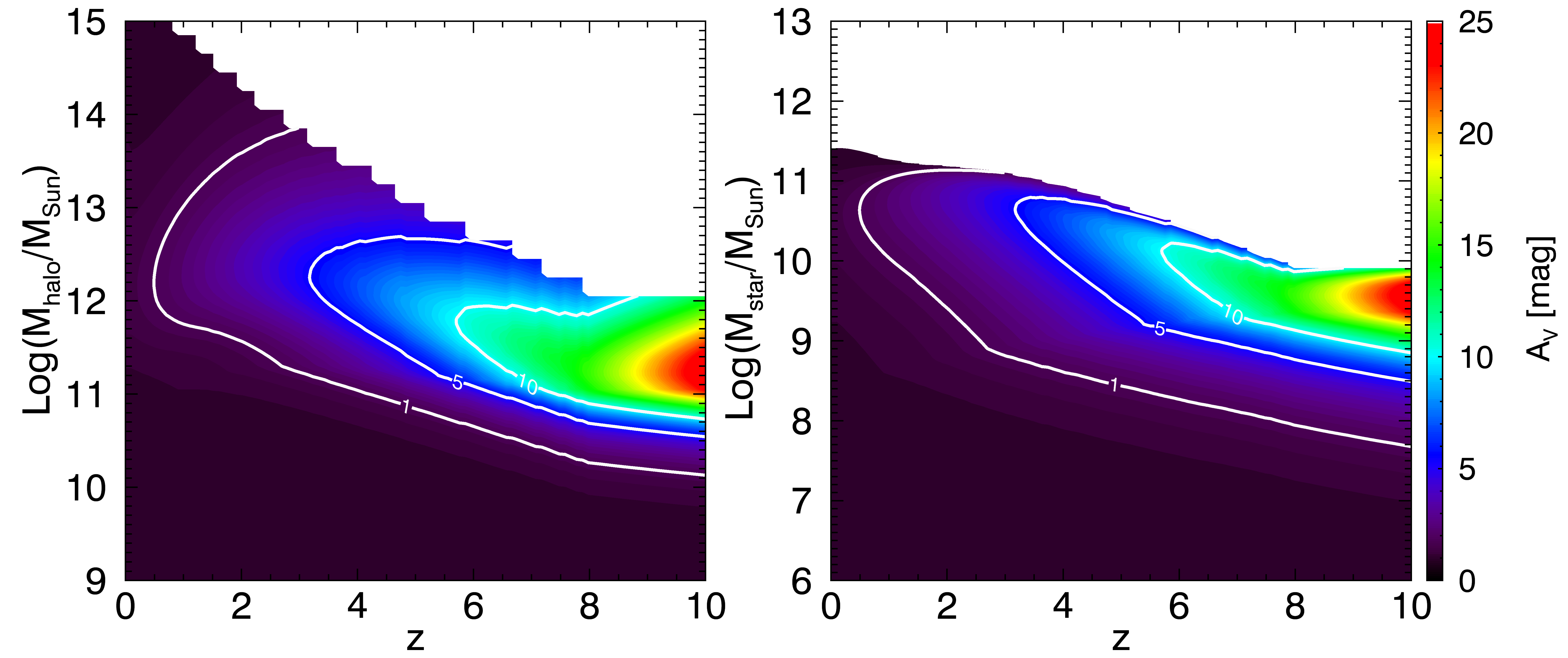}
\caption{Extinction due to dust, in units of visual magnitudes.  Left panel: \av~ as a function of halo mass and redshift.  
Right panel: \av~as a function of stellar mass and redshift.  Contour lines at 1, 5, and 10 mag are overlaid for clarity.  \label{fig:mass-z}}
\end{figure*}

In Figure \ref{fig:mdust_multi}, we again show \mdust~as a function of \mstar, this time overplotting observations from the literature.
In the first panel, we overplot observations from \citet{Remy-Ruyer_2015} and the \emph{Herschel} Reference Survey \citep[HRS;][]{Ciesla_2014, Boselli_2015}, and we find good agreement between our model and observed dust masses at $z=0$, for stellar masses ranging from about $10^7$ to $10^{11}~\msun$. 
At $z=1$ and $z=2$, the dust masses predicted by our model are in good agreement with the observations by \citet{Santini_2014}. 
We underpredict the observations by \citet{daCunha_2015} at these $z=1$ and 2 by roughly $0.5$ to $1$ dex.  This is not too surprising, however, since the \citet{daCunha_2015} observations are of sub-millimeter galaxies (SMGs), which have higher than typical dust infrared luminosities ($>10^{12}L_\odot$), which are driven by high SFRs in excess of 100 $\msun\rm{yr}^{-1}$.     Sub-millimeter selection basically selects for SFR, and the \citet{daCunha_2015} SMGs have SFRs about a factor of 3 higher than main sequence galaxies of the same stellar mass.  Galaxies such as those in the da Cunha et al. sample, which includes some starbursts, may have quickly enriched the ISM with metals and dust on timescales shorter than that for main sequence galaxies and are not accounted for by the models adopted in this study.

On the low-mass end of galaxies  ($\mstar<10^{10}$), it will be interesting to see how well our model reproduces the \mdust-\mstar~trend at redshifts $z\ge 1$.  However, as we show in \S \ref{sec:flux}, detections of the dust emission in large samples of low-mass, high-redshift galaxies would be a challenging prospect for observing programs with present-day telescopes.

We find that the relation between galaxy dust mass and stellar mass can be parameterized as a broken power law, 
\begin{equation}\label{eq:power}
 \log\left( \frac{\mdust}{\mdust_{,0}}  \right) = 
  \begin{cases} 
    \alpha_1\log \left( \frac{\mstar}{\mstar_{,0}} \right)  &  \text{if } \mstar \le \mstar_{,0} \\
    \alpha_2\log \left( \frac{\mstar}{\mstar_{,0}} \right)  &  \text{if } \mstar > \mstar_{,0} ,
  \end{cases}
\end{equation}
where $\mdust_{,0}$ is the zero point of the dust mass, and  $\alpha_{1,2}$  are the slopes below and above $\mstar_{,0}$, the stellar mass at a given redshift where the break in the power law occurs. The break point in the \mdust-\mstar~relation results from the prescription we use for the dust-to-gas ratio, equation \citep[\ref{eq:dgr};][]{Remy-Ruyer_2014}, also a broken power law, which depends on \mstar~and the SFR via the metallicity.  Both $\mdust_{,0}$ and $\mstar_{,0}$ are functions of redshift.

The amount of dust in a galaxy is determined by the amount of available metals and gas, both of which are linked to star formation activity.  Recent observational studies have shown that the dust-to-gas ratio of nearby galaxies may be characterized as a function of the gas-phase metallicity \citep{Remy-Ruyer_2014}.  Theoretical work by \citet{Popping_2017}, who use semi-analytical models to follow the production of interstellar dust, reproduces this observed trend and suggests that it is driven by the accretion of metals onto dust grains and the density of cold gas.  Like \citet{Popping_2017}, we find that the normalization of the dust-mass-stellar-mass relation, $\mdust_{,0}$,  increases from $z=0$ to $z=9.5$.

We perform least-squares fits to the predicted curves in Figures \ref{fig:mdust} and \ref{fig:mdust_multi} to determine the parameters in equation (\ref{eq:power}).  The results for $z=0$ to $z=9.5$ are summarized as follows:
\begin{equation}\label{eq:dusfit}
\begin{split}
\alpha_1 &= 1.20\pm 0.02 \\
\alpha_2 &= 0.75\pm 0.02 \\
\log \mdust_{,0} &= (6.0 \pm 0.1) + (1.8\pm 0.1)\log(1+z) \\
\log \mstar_{,0} &= (8.4 \pm 0.1) + (1.0\pm 0.2)\log(1+z).  \\
\end{split}
\end{equation}

\vspace{1cm}

\subsection{Evolution of Dust Optical Depth}\label{sec:tau0}
In Figure \ref{fig:mass-z} we present contour maps of the optical depth due to dust in terms of visual extinction, $\av = 1.086\tau_V$ \citep[e.g.,][]{Draine_2011}, calculated in the rest frames of the galaxies.  The left panel of Figure \ref{fig:mass-z} displays the redshift evolution of \av~in terms of \mhalo, and the right panel displays the evolution of \av~in terms of \mstar. 
The maps demonstrate that for constant values of \mhalo~or \mstar, \av~increases with increasing $z$.  For instance, while a $10^{12}~\msun$ halo mass galaxy has an extinction due to dust of $\av\approx 0.8$ mag at $z=0$, a similar galaxy at $z=4$ has an extinction of $\av\approx 7$ mag.  This basic trend holds for the optical depth at other wavelengths, with the galaxies having overall higher optical depths at shorter wavelengths and lower optical depths at longer wavelengths. 

It is possible that we slightly overpredict \av~at high redshifts.  As we discuss in further detail in \S\ref{sec:caveats}, our approximation of spherical symmetry could lead to overestimates of the optical depth, particularly for massive galaxies, in which the bulk of interstellar dust is typically observed to reside in the disk.  The rise in \av~at high redshift is mostly driven by \rdust, since $\tau_\nu\propto 1/\rdust^2$ (equation \ref{eq:tau1}), and since $\rdust\propto (1+z)^{-1}$ (equations \ref{eq:rgal} and \ref{eq:rvir}; Figure \ref{fig:model}).  Thus, with our assumption of spherical geometry, and given that we have defined the optical depth as the integrated value through to the galactic center (equation \ref{eq:tau0}), the values we have derived for \av, are most likely upper limits.

\begin{figure*}[!]
\centering
\includegraphics[width=7in]{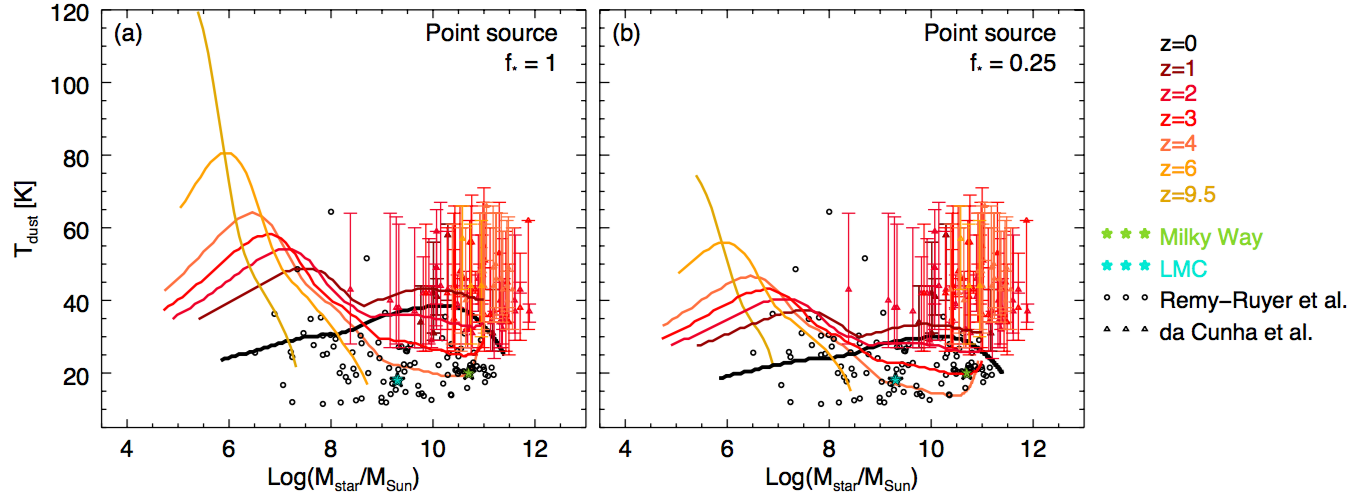}
\caption{Dust temperature as a function of stellar mass and redshift for the point source model. 
The fraction of starlight absorbed by dust is $f_\star = 1$ and $0.25$ in panels (a) and (b).  
The overplotted symbols represent observed values from the literature. \label{fig:tdust1}}
\end{figure*}

\begin{figure*}[!]
\centering
\includegraphics[width=7in]{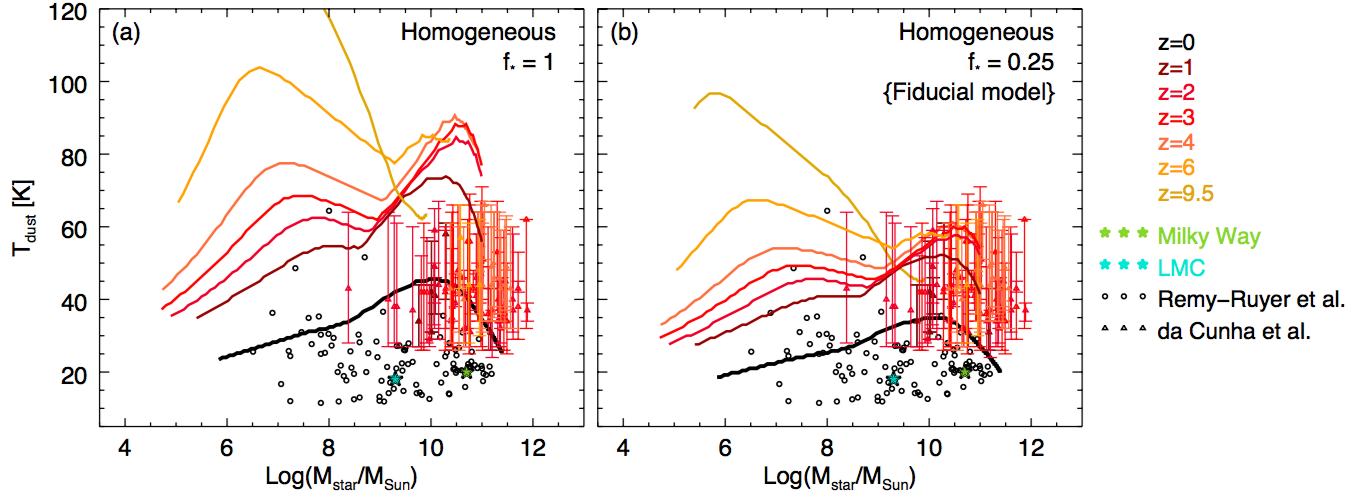}
\caption{Dust temperature as a function of stellar mass and redshift for the homogeneous model. 
The fraction of starlight absorbed by dust is $f_\star = 1$ and $0.25$ in panels (a) and (b).  
The overplotted symbols represent observed values from the literature. In the left-hand plot, \tdust~reaches a maximum of 173.6 K at Log$(\mstar/\msun)=5.8$.
 \label{fig:tdust2}}
\end{figure*}

\subsection{Evolution of Dust Temperature}
We use our results for \mdust~and $\tau_\nu$ in the previous sections to determine \tdust~from equation (\ref{eq:tdust}).  
In Figures  \ref{fig:tdust1} and \ref{fig:tdust2}, we present predictions for the evolution of galaxy dust temperature as a function of stellar mass.  We show results for the two galaxy geometries we consider here, the ``point source'' and ``homogeneous'' models illustrated in Figure \ref{fig:geometry}.  We also show results for two different surface area covering fractions of dust, $f_\star=1$ and $f_\star=0.25$.  Both models show that the \tdust-\mstar~relation is not monotonic, but rather peaks at characteristic values of \mstar.  In both figures, we overplot data points of measured galaxy dust temperatures from the literature.

The point source model (Figures  \ref{fig:tdust1}) shows that for most galaxy stellar masses ($\mstar\gtrsim 10^{6.5}$ \msun), \tdust~tends to increase over time, and is a poor representation of the observations.  Whereas observations suggest that dust temperatures in higher mass galaxies tend to get cooler with time, the point source model predicts the opposite.  By contrast, the model in which stars and the ISM are homogeneously distributed (Figures  \ref{fig:tdust2}) suggests that \tdust~tends to decrease with time.  At all redshifts,  our homogeneous, $f_\star=0.25$ model is in better agreement with the observations by \citet{Remy-Ruyer_2015} and \citet{daCunha_2015} than the $f_\star=1$ model.  We take the former to be our fiducial model; Table \ref{table2} lists the values of \tdust~for a $10^{12}$ \msun~halo mass galaxy in this model.  That the $f_\star=0.25$ model is in better agreement with the observations than the $f_\star=1$ model is not surprising.  
This implies that galactic dust does not have a covering fraction of 100\% and that there are regions in any given galaxy where starlight escapes without being absorbed by dust.  The actual covering fraction is almost certain to vary widely from galaxy to galaxy.

Focusing now on Figure \ref{fig:tdust2}, starting at about $z=6$, the dust temperature tends to cool down for galaxies of fixed stellar masses,  as the Universe evolves. 
At each redshift, there are two peaks in the \tdust-\mstar~relation.  
As the redshift decreases, the first peak shifts towards higher and higher values of \mstar.  
For instance, between $z=6$ and $z=3$, the stellar mass at which \tdust~peaks shifts from $\mstar=10^{6.6}$ to $10^{7.7}$ \msun.  
The peak near the high-mass end is more stable and does not display a similar systematic shift over time.

Our model predicts a population of high-redshift ($z\gtrsim 2$), low-mass galaxies ($\mstar\approx 10^6$ to $10^{8.5}$ \msun) with fairly hot dust.  
From $z=6$ to $z=4$, these galaxies have dust that is around the same temperature as---if not hotter than---the dust in their more massive counterparts at the same redshifts.  Unfortunately, there are no observational constraints on the star formation history of such galaxies, and so with the current state of knowledge, indirect methods would have to be used to infer the robustness of the peak temperatures our model predicts.

\begin{figure}[t]
\epsscale{1.}
\plotone{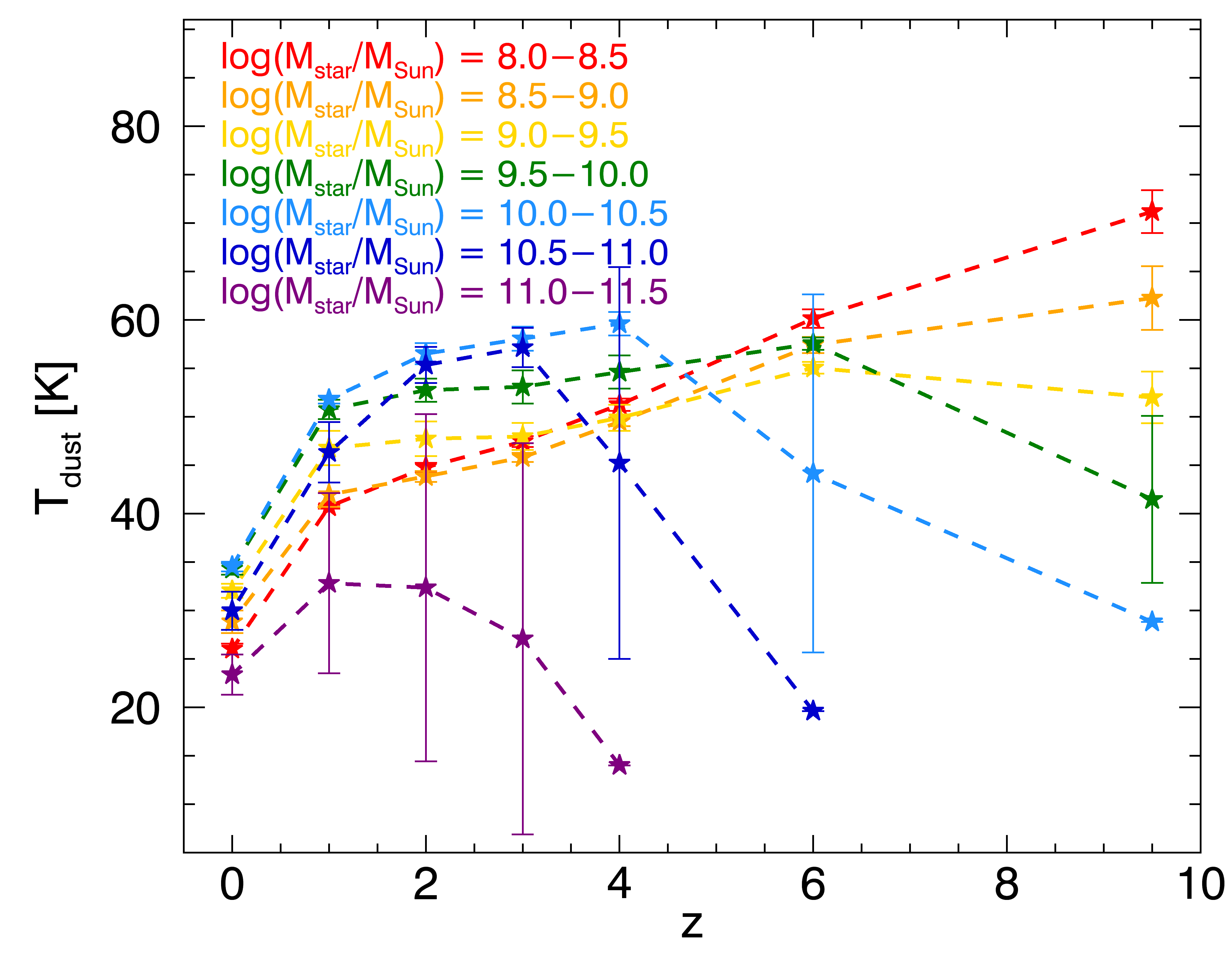}
\caption{Evolution of mean dust temperature for fixed stellar masses.  The error bars represent the $1\sigma$ scatter about the mean temperature for the given range of stellar masses.}
 \label{fig:tdustz}
\end{figure}

In Figure \ref{fig:tdustz} we plot dust temperatures as a function of redshift for fixed stellar masses.  We find that dust temperatures of galaxies of all stellar masses evolve markedly with redshift.  Starting the present era to $z=6$, galaxies having stellar masses from $10^8$ to $10^{10}$ \msun~have dust temperatures which increase monotonically from about $25$-$35$ to $\sim 55$--60.  Higher mass galaxies first display increases in \tdust, followed by decreases in the dust temperature as they evolve to higher redshifts. 

\citet{Viero_2013} stacked \emph{Herschel} images of a stellar mass selected sample of galaxies, and \citet{Bethermin_2015} performed a similar stacking analysis of  \emph{Spitzer}, \emph{Herschel}, LABOCA, and AzTEC data for galaxies taken from the COSMOS field.  These authors find that the dust temperature tends to increase with redshift, up to $z=6$, for galaxies of all stellar masses.  While we find a similar trend for galaxies having stellar masses $\mstar<10^{10}\msun$, our results are at odds with the observations for massive galaxies.   That the massive galaxies in our model do not show a steady increase in \tdust~with redshift is primarily a reflection of our geometric model for the ISM.  As discussed in Section \ref{sec:tau0}, our approximation of spherical symmetry may result in overestimates of $\tau_\nu$, especially for high-redshift, massive galaxies, where most interstellar dust is observed in the disk.   Over-predicting $\tau_\nu$ for high-mass galaxies naturally leads to the non-monotonic evolution of $\tdust$ observed in Figure \ref{fig:tdustz}.  Nevertheless, the values we derive for \tdust~at any fixed redshift are in general agreement with the observational results of \citet{Viero_2013} and \citet{Bethermin_2015}, and with the theoretical models of \citet{Cowley_2017}.  At any given redshift, we predict temperatures a factor of roughly 1.5 higher than these authors.  This systematic offset may partly be due to the choice of modified blackbody models fitted by these authors and partly a result of how they selected sources.  Indeed, other authors who performed stacking analyses but fitted different models or had different selection criteria, including \citet{Pascale_2009}, \citet{Amblard_2010}, and \citet{Elbaz_2010}, report higher dust temperatures in alignment with our results.  We note that these latter two studies, which are based on galaxy samples selected by \emph{Herschel}, may be biased in temperature.  Due to the varying sensitivity of the PACS instrument with wavelength, hotter galaxies are easier to detect.

\begin{figure}[t]
\epsscale{1.}
\plotone{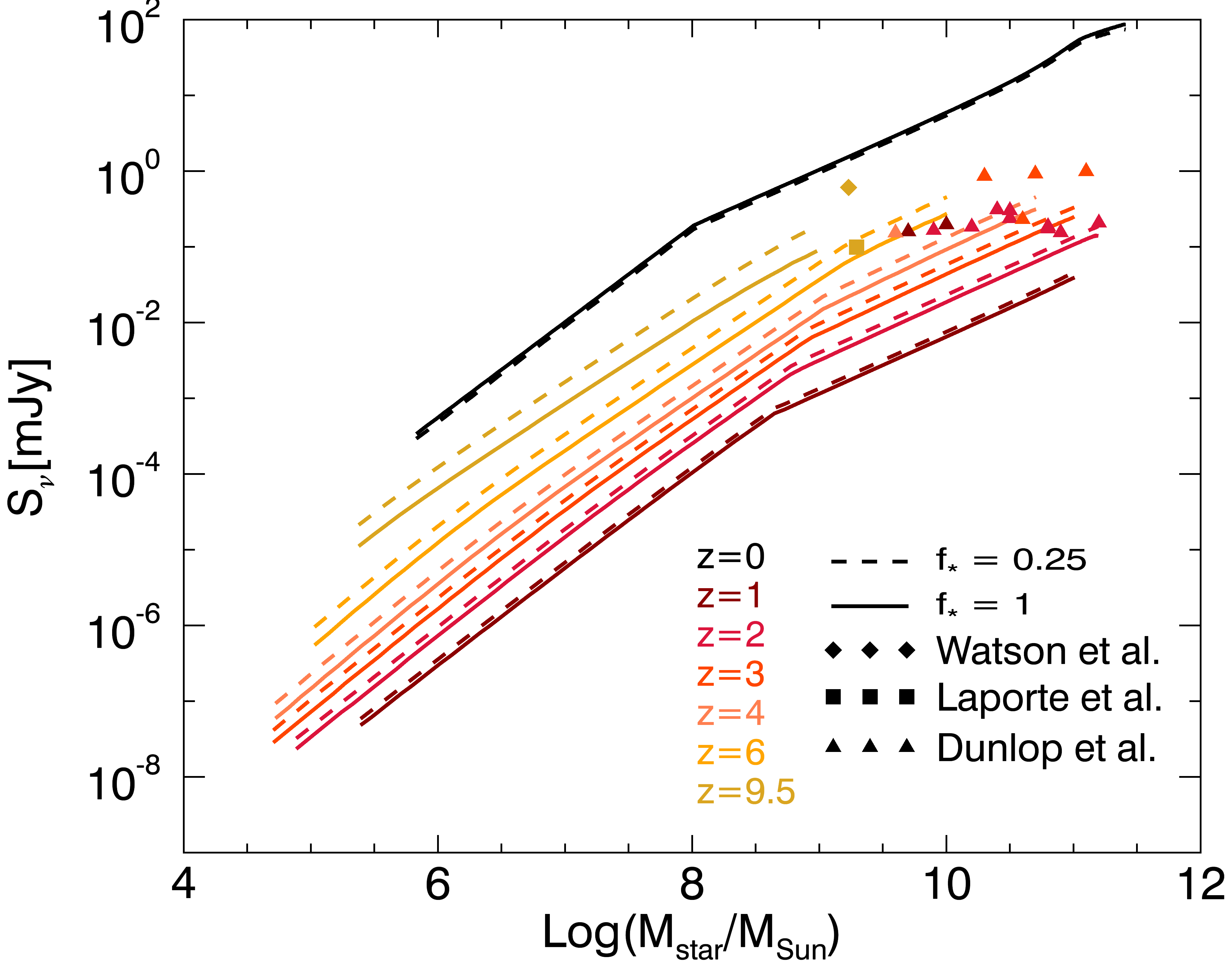}
\caption{Flux density at 1.1 mm as a function of stellar mass for $z=0$--9.5. \label{fig:snu1}}
\end{figure}

\subsection{Predictions for the Observable Flux Density}\label{sec:flux}
In light of recent ALMA observations that have detected large amounts of dust in galaxies during the epoch of reionization \citep[e.g.,][]{Watson_2015, Laporte_2017}, we are motivated to make predictions of the flux density due to dust in galaxies.  We assume that the dust in a galaxy, of total mass \mdust, will rise to an average temperature $\tdust$, due to heating by starlight, and the dust will re-emit most of the light in the infrared.  The galaxy will have a flux density of 
\begin{equation}
S_\nu = \frac{\kappa_\nu B_\nu(\tdust) \mdust (1+z)}{d_L^2},
\end{equation}
where $B_\nu(\tdust)$ is the Planck function, and $d_L$ is the luminosity distance to the galaxy.  The emission is assumed to be optically thin.

\begin{table*}[t]
\centering
\caption{Observations plotted in Figures \ref{fig:snu1} and  \ref{fig:snu2}.}\label{table3}
\begin{threeparttable}
\setlength{\tabcolsep}{12pt}
\begin{tabular}{lccccc}
\tableline\tableline
Galaxy name  &   Redshift    &  \snu  & \mstar         &   SFR$_{\rm IR}$   & Reference \\
             &               &  (mJy) & ($10^9$ \msun) &  (\msun~yr$^{-1}$) &  \\
\tableline
A1689-zD1    & $7.5\pm 0.2$  &  $0.61\pm 0.12$ & $1.7^{+0.7}_{-0.5}$  &  $9^{+4}_{-2}$  & 1  \\
A2744\_YD4   & $8.38^{+0.13}_{-0.11}$   &  $0.099\pm 0.023$ & $1.97^{+1.45}_{-0.66}$  &  $20.4^{+17.6}_{-9.5}$  & 2  \\
UDF1  &  3.00  &  $0.924 \pm 0.076$   & $50^{+13}_{-10}   $ & $326 \pm 83 $  &  3 \\
UDF2  &  2.79  &  $0.996 \pm 0.087$   & $126^{+52}_{-37}  $  & $247 \pm 76$  &  3 \\  
UDF3  &  2.54  &  $0.863 \pm 0.084$   & $20 ^{+8}_{-6} $  & $195 \pm 69$  &  3 \\  
UDF4  &  2.43  &  $0.303 \pm 0.046$   & $32^{+13}_{-9}  $  & $94 \pm 4$  &  3 \\  
UDF5  &  1.76  &  $0.311 \pm 0.049$   & $25^{+10}_{-7}  $  & $102 \pm 7$  &  3 \\  
UDF6  &  1.41  &  $0.239 \pm 0.049$   & $32^{+8}_{-7}  $  & $87 \pm 11$  &  3 \\   
UDF7  &  2.59  &  $0.231 \pm 0.048$   & $40^{+10}_{-8}  $  & $56 \pm 22$  &  3 \\   
UDF8  &  1.55  &  $0.208 \pm 0.046$   & $159^{+65}_{-46}  $  & $149 \pm 90$  &  3 \\  
UDF9  &  0.67  &  $0.198 \pm 0.039$   & $10^{+3}_{-2}  $ & $23 \pm 25$  &  3 \\  
UDF10  & 2.09  &  $0.184 \pm 0.046$   & $16^{+7}_{-5}  $  & $45 \pm 22$  &  3 \\  
UDF11  & 2.00  &  $0.186 \pm 0.046$   & $6^{+16}_{-13} $  & $162 \pm 94$ &  3 \\ 
UDF12  & 5.00  &  $0.154 \pm 0.040$   & $4^{+2}_{-1}  $  & $37 \pm 14$  &  3 \\   
UDF13  & 2.50  &  $0.174 \pm 0.045$   & $63^{+16}_{-13}  $  & $68 \pm 18$  &  3 \\   
UDF14  & 0.77  &  $0.160 \pm 0.044$   & $5^{+1}_{-1}  $  & $44 \pm 17$  &  3 \\   
UDF15  & 1.72  &  $0.166 \pm 0.046$   & $8^{+3}_{-2}  $  & $38 \pm 27$  &  3 \\
UDF16  & 1.31  &  $0.155 \pm 0.044$   & $79^{+21}_{-16}  $  & $40 \pm 18$  &  3 \\

\tableline
\end{tabular}
\textbf{References.} (1) Watson et al. (2015); (2) Laporte et al. (2017); (3) Dunlop et al. (2017).
\end{threeparttable}
\end{table*}

Figure \ref{fig:snu1} presents the galaxy flux density at an observed wavelength of $1.1$ mm, as a function of \mstar~for $z=0$ to $z=9.5$.  For fixed redshifts, $S_\nu$ increases with galaxy stellar mass.  
For fixed stellar masses, for sources at redshifts $z\gtrsim 1$, $S_\nu$ decreases with time.  
This trend is due to the combination of two effects: (1) the negative $K$-correction \citep[e.g.,][]{Hughes_1998, Barger_1999, Blain_2002, Lagache_2005}; and (2) our fiducial model predicts that \tdust~increases with $z$.  At redshifts $z>1$, the far-infrared radiation emitted in distant galaxies is redshifted to sub-mm wavelengths, and the resulting negative $K$-correction counteracts the dimming of galaxies caused by their cosmological distances.  
If more distant galaxies have hotter dust, then the observed flux from these galaxies originated from radiation emitted closer to the peak of the black body radiation curve than nearby galaxies.  For instance, photons emitted from dust in a galaxy at $z=9.5$ had rest wavelengths of $\lambda_0=116$ \micron, while photons emitted from a source at $z=2$ had $\lambda_0=550$ \micron.  From Figure \ref{fig:tdust2}, one can see that the dust temperature of a $10^8$ \msun~stellar mass galaxy at $z=9.5$ is $\sim 74$ K, corresponding to a peak in the black body radiation curve at 39 \micron.  A $10^8$ \msun~galaxy at $z=2$ has $\tdust\approx 45$ K, corresponding to 64 \micron.  Thus, the observed flux at 1.1 mm from the more distant galaxy at $z=9.5$ originated from dust whose emission was closer to the peak of the black body radiation curve, compared to the galaxy at $z=2$. 

\begin{figure}[h]
\epsscale{1.}
\plotone{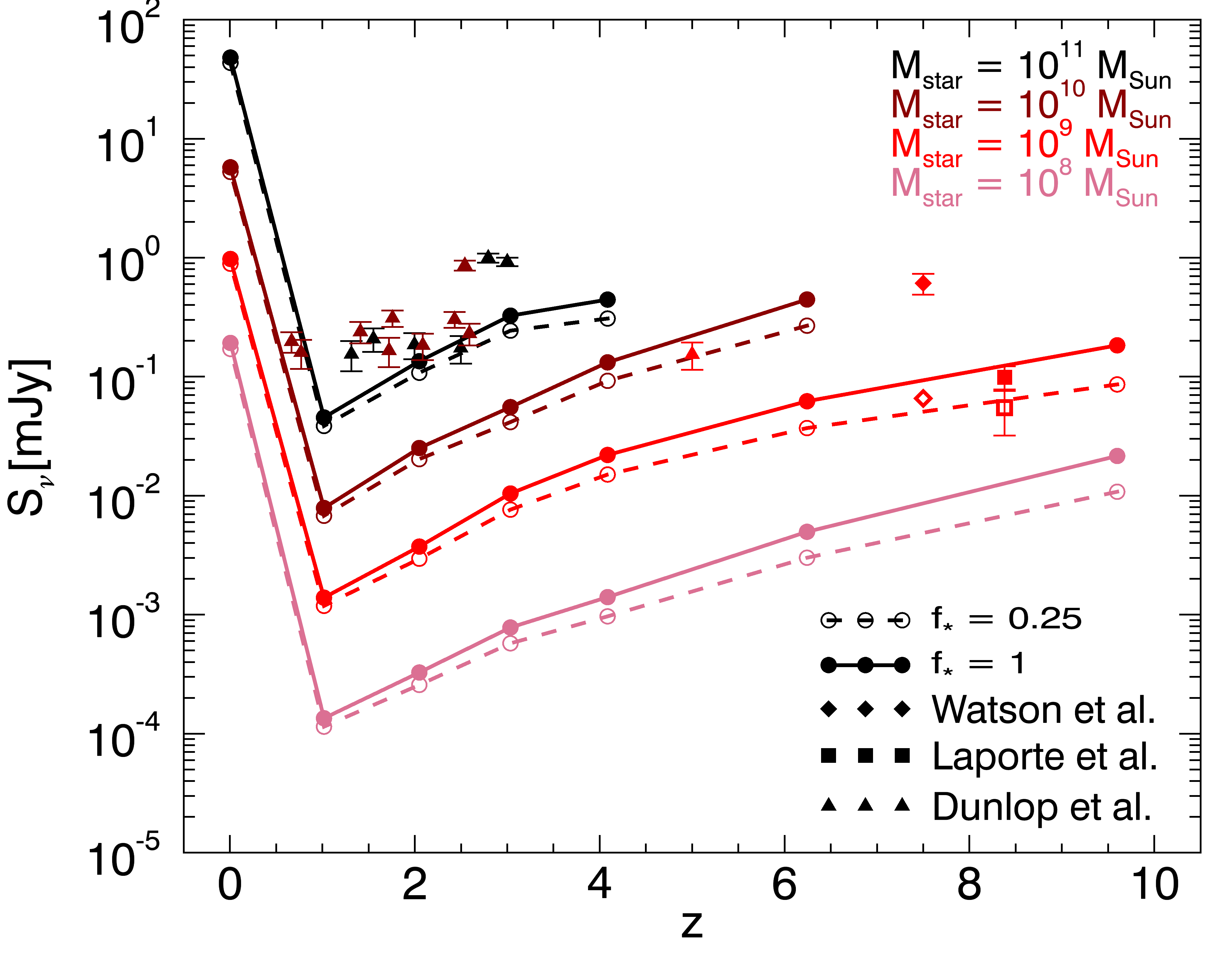}
\caption{Flux density at 1.1 mm as a function of redshift, for fixed stellar masses: $10^8$, $10^9$, $10^{10}$, and $10^{11}$ \msun.  The values of the overplotted observations are provided in Table \ref{table3}.   We also overplot the fluxes measured by  Watson et al. (2015) and Laporte et al. (2017), corrected for magnification, with open diamond and square symbols, respectively. \label{fig:snu2}}
\end{figure}

In Figure \ref{fig:snu2}, we show results in terms of $S_\nu$ as a function of $z$, for fixed stellar masses: $\mstar=10^8$, $10^9$, and $10^{10}$ \msun.  For a given stellar mass, $S_\nu$ first decreases with time, and then it rises sharply from $z=1$ to $z=0$.  We overplot the observed $\sim 1$ mm continuum fluxes of galaxies recently observed with ALMA.  \citet{Watson_2015} observed the lensed galaxy A1689-zD1 between $1.2$ and $1.4$ mm and detected a flux of $\snu=0.61\pm 0.12$ mJy.  Located at $z\approx 7.5$ and magnified by a factor of 9.3, A1689-zD1 has a stellar mass, dust mass, and SFR of $\mstar\approx 2\times 10^9$ \msun, $\mdust\approx 4\times 10^7$ \msun, and $\rm{SFR}\approx 9$ $\msun\rm{yr}^{-1}$.  \citet{Laporte_2017} observed the lensed galaxy A2744\_YD4 at 0.84 mm.  At a redshift of about 8.4 and magnified by a factor of $\sim 1.8$,  A2744\_YD4 has a stellar mass, dust mass, and SFR of $\mstar\approx 2\times 10^9$ \msun, $\mdust\approx 6\times 10^6$ \msun, and $\rm{SFR}\approx 20$ $\msun\rm{yr}^{-1}$.  Dunlop et al. (2017) conducted the first, deep ALMA image of the \emph{Hubble} Ultra Deep Field, detecting 16 sources at 1.3 mm.  The sources have high stellar masses, with 13 out of 16 having $\mstar>10^{10}$ \msun.  Fifteen of the sources are located at redshifts $0.7\le z\le 3$, and one source is located at $z=5$.  The observed fluxes and other properties of all the sources plotted in Figures \ref{fig:snu1} and \ref{fig:snu2} are summarized in Table \ref{table3}.

To date, observations of the dust emission in sources at $z\gtrsim 1$ have been restricted to galaxies having stellar masses $\gtrsim 10^9$ \msun.  To achieve the sensitivities necessary to detect the dust emission of single, high-redshift $L^\star$ galaxies at these masses requires total observing times of $\sim 2$--3 hours with ALMA \citep[e.g.,][]{Laporte_2017}.  Since typical lower mass galaxies ($\mstar<10^9$ \msun) are intrinsically fainter, the much longer observing times needed to detect their dust emission at $z>1$ may be prohibitive.  
Yet serendipitous occurrences, such as gravitational lensing, could possibly aid in the detection and characterization of the low-mass, star-forming population of galaxies in the early Universe.  If and when low-mass galaxies begin to be detected in large numbers, it may be easier (e.g., less time-consuming) to first detect galaxies at higher redshifts, since according to our model, their observed millimeter flux is expected to exceed that of lower redshift galaxies by $\sim 1$ to 2 orders of magnitude.

\begin{figure*}[t]
\centering
\includegraphics[width=6in]{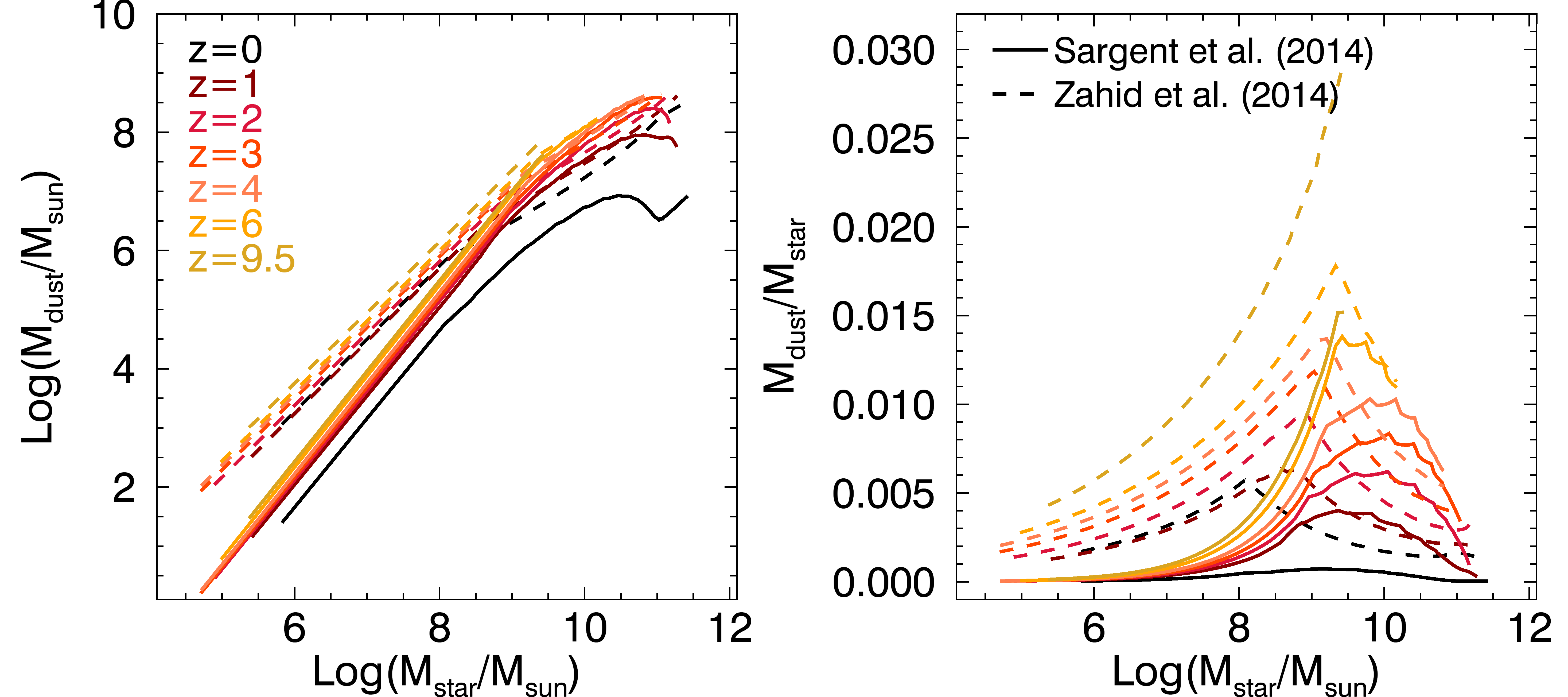}
\caption{Galaxy dust mass versus stellar mass for different redshifts. \mdust~is calculated using the Sargent et al. (2014; solid lines) and Zahid et al. (2014; dashed lines) prescriptions for \mgas.   \label{fig:mdust2}}
\end{figure*}

\section{Caveats \& Limitations}\label{sec:caveats}
We now discuss in further detail some of the key assumptions of our model and assess their impact on the results.

\subsection{Galactic geometry and radial distribution of dust}
In \S\ref{sec:model} we model galaxies as spherically symmetric with either most of the dust concentrated around the nucleus or else homogeneously mixed with gas and stars.  In reality, dust is often consolidated in the disk of large galaxies, and so the assumption of spherical symmetry may result in overestimates of the optical depth for these systems. 

In equation (\ref{eq:tau0}), we assume that galactic dust has a simple power-law density distribution, with $\gamma=0$.  The assumption that the dust profile is constant with radius may seem unfounded, given observations that dust content varies with galactic radius \citep[e.g.,][]{Boissier_2005, Munoz-Mateos_2009}.  
Yet appropriate values for $\gamma$ and \rdust, about which we are equally ignorant, are certain to vary significantly from galaxy to galaxy.  Since there are infinite combinations of $\gamma$ and $\rdust$ that produce identical values of $\tau_\nu$---that is, since $\gamma$ and $\rdust$ are essentially degenerate---we decide to absorb our ignorance about both quantities into our definition of \rdust~in equation (\ref{eq:tau0}), where we let $\rdust=\rgal$, (recalling that \rgal~is the half-mass galactic radius).  For example, for two galaxies with identical values of $\kappa_\nu$ and \mdust, $\tau_\nu$ for one galaxy with $\rdust=\rgal$ and $\gamma=0$ is equivalent to $\tau_\nu$ for the second galaxy with $\gamma=1/3$ and $\rdust=2\rgal$.  

\begin{figure*}[t]
\centering
\includegraphics[width=6in]{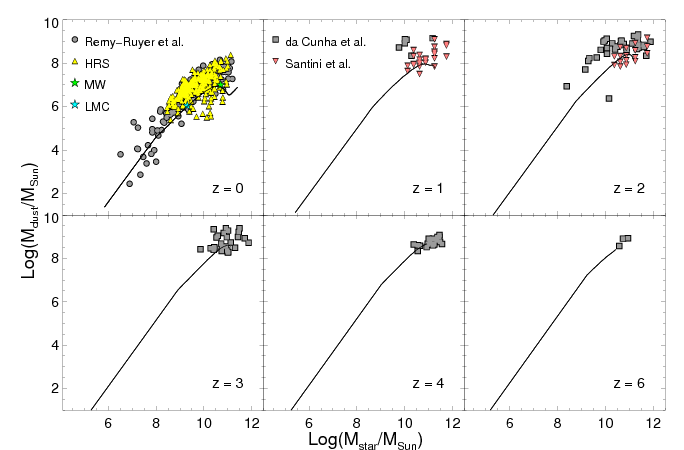}
\caption{Galaxy dust mass versus stellar mass for different redshifts.  The plots are analogous to those in Figure \ref{fig:mdust_multi2}, accept here, \mdust~is calculated using the Sargent et al. (2014) prescription for \mgas.  The green and turquoise stars represent values for the Milky Way and the Large Magellanic Cloud, respectively.  The triangles are measurements from the Herschel Reference Survey.   \label{fig:mdust_multi2}}
\end{figure*}

Another source of uncertainty in our model is the choice of spin parameter, $\lambda$, which is expected to link galaxy disk size and halo size (see equation \ref{eq:rgal}), under the assumption that the collapsing baryonic matter of a galaxy inherits the same specific angular momentum as the halo \citep{Fall_1980}.  In \S\ref{sec:radius}, we assume that the half-mass radius and dark matter halo radius, for all galaxy masses at all redshifts, are linked by a single value, $\rgal/R_{\rm halo}=0.018$, corresponding to $\lambda = 0.036$ \citep{Somerville_2018}.  Some simulations suggest that there is significant scatter---about two orders of magnitude---about the mean value of $\lambda$ \citep{Teklu_2015, Zavala_2016}, and that this scatter depends in part on galaxy morphology \citep[e.g.,][]{Teklu_2015}.  The results of \citet{Somerville_2018}, who demonstrate that $\lambda$ is roughly independent mass and  weakly evolves with redshift, are in general agreement with other recent studies of the relationship between galaxy and halo size \citep[e.g.,][]{Shibuya_2015, Kawamata_2015, Huang_2017}.  Yet if our adopted value of $\lambda$ leads to under- or overestimates of \rgal~for galaxies at certain masses and epochs, these inaccuracies will naturally propagate into our estimates of \rdust~and $\tau_\nu$.

\subsection{Evolution of gas mass}
We use a prescription for $\mgas(\mstar,z)$ determined by \citet{Zahid_2014}, who determined a relation between metallicity and stellar-to-gas mass ratio for galaxies at $z\lesssim 1.6$ and with $\mstar\gtrsim 10^9~\msun$.  We extrapolate this relation for higher redshifts and lower stellar masses.  If the interstellar environments of low-mass galaxies manifest in significantly different relationships between $\mathcal{Z}$, \mstar, and \mgas, and if ISM conditions evolve with redshift, then the Zahid et al. relation may break down in unexpected ways in the low-\mstar, high-$z$ regimes.  Further observational tests are required to confirm or rule out the universal metallicity relation upon which equation (\ref{eq:mgas}) is based.  Nevertheless, it is promising that the Zahid et al. relation is consistent with that of \citet{Andrews_2013}, who measure the MZ relation down to $\mstar\approx 10^{7.5}~\msun$. 

The predicted dust masses are sensitive to the functional form of \mgas, (equation \ref{eq:mdust}).  To give a sense of how the evolution of \mgas~affects \mdust, we present additional calculations for \mdust, using an alternative prescription for \mgas.  \citet{Sargent_2014} compiled a sample of 131 massive ($\mstar >10^{10}$ \msun), star-forming galaxies at redshifts $z\lesssim 3$.  They derived a Schmidt-Kennicutt relation:
\begin{equation}\label{eq:sargent}
\begin{split}
\log \left(\frac{M_{\rm mol}}{\msun}\right) &= (9.22\pm 0.02) \\
&+ (0.81\pm 0.03)\log\left(\frac{\rm{SFR}}{\msun~\rm{yr}^{-1}}\right),
\end{split}
\end{equation}
where $M_{\rm mol}$ is the galactic molecular mass.  Equation (\ref{eq:sargent}) is appealing as a comparison to the \citet{Zahid_2014} formulation for \mgas, because it is directly applicable for galaxies at higher redshifts.  We do not attempt to estimate the total gas mass from equation \ref{eq:sargent} but calculate the dust mass using $\mgas = M_{\rm mol}\times\rm{DGR}$.
Using the \citet{Sargent_2014} formulation for the molecular mass of galaxies, in Figures \ref{fig:mdust2} and \ref{fig:mdust_multi2} we display plots of \mdust~as a function of \mstar, analogous to Figures \ref{fig:mdust} and \ref{fig:mdust_multi}.  Figure  \ref{fig:mdust2} shows that while the \citet{Zahid_2014} formulation for \mgas~results in higher predictions for \mdust~for most stellar masses, for galaxies with $\mstar\gtrsim 10^{9.5}~\msun$, the Zahid et al. and Sargent et al. formulations come into better agreement.  Not too surprisingly, using equation (\ref{eq:sargent}) leads to a model for $\mdust$ that underpredicts observed values in high-mass galaxies (Figure \ref{fig:mdust_multi2}).  This is because we did not account for the total galactic gas mass here, only $M_{\rm mol}$.  However, \mdust~calculated using equation (\ref{eq:sargent}) is in good agreement with low-mass galaxies with $\mstar\lesssim 10^8~\msun$.  Moreover, for redshifts $z\gtrsim 1$, \mdust~calculated using equation (\ref{eq:sargent}) comes into better agreement with the observations of high-mass galaxies, since the gas fraction in galaxies is expected to be increasingly dominated by molecular gas as redshift increases.

The total gas mass fraction,  defined  $f_{g,\rm{tot}}=\mgas/(\mgas + \mstar)$, is the subject of a number of studies.  For stellar masses in the range $10^{10}$--$4\times 10^{11}\msun$, the \citet{Zahid_2014} relation predicts $f_{g,\rm{tot}}\approx 0.28$, 0.40, 0.48, and 0.55 for redshifts $z=1$, 2, 3, and 4, respectively, though the authors caution the extrapolation of their relation beyond $z=1.6$ (private communication).  
These values are quite consistent with the \emph{molecular} gas fractions, $f_{g,\rm{mol}}=M_{\rm mol}/(M_{\rm mol} + \mstar)$, derived in many studies from CO and dust observations of galaxies having a similar range of stellar masses.  \citet{Daddi_2010} measured $f_{g,\rm{mol}}\approx 0.6$ for 6 galaxies with $\mstar=0.33$--$1.1\times 10^{11}\msun$ at $z=1.5$.  \citet{Tacconi_2010} studied 19 galaxies with $\mstar=0.3$--$3.4\times 10^{11}\msun$.  They measured  $f_{g,\rm{mol}}=0.2$--$0.5$ at $z\approx 1.1$ and $f_{g,\rm{mol}}=0.3$--$0.8$ at $z\approx 2.3$.   Later on, \citet{Tacconi_2013} found similar results with a larger sample of 52 galaxies, measuring average molecular gas fractions of 0.33 and 0.47 at $z\sim 1.2$ and 2.2.  \citet{Magdis_2012} measured $f_{g,\rm{mol}}\approx 0.36$ for a $\mstar=2\times 10^{11}\msun$ galaxy at $z=3.21$.  \citet{Saintonge_2013} measured $0.45$ for $\mstar\sim 10^{10}\msun$ galaxies at $z=2.8$.  More recently, \citet{Bethermin_2015} used observations of dust emission in massive ($\sim 6\times 10^{10}\msun$) galaxies to measure $f_{g,\rm{mol}}=0.16$--0.35 at $z<1$, 0.27--0.41 at $1<z<2$, $\sim 0.5$ at $2<z<3$, and $\sim 0.6$ at $3<z<4$.  And in an extensive study of 145 galaxies from the COSMOS survey, \citet{Scoville_2016} measured molecular gas fractions of $0.16$--$0.67$ at $z\approx 1.5$, $0.24$--$0.75$ at $z\approx 2.2$ and $0.23$--$0.85$ at $z\approx 4.4$.

While it is generally agreed that high-redshift galaxies are gas-dominated, the details of the redshift evolution of $f_{g,\rm{mol}}$ are uncertain.   By re-expressing the gas fraction (total or molecular) as $1/[1+(t_{\rm dep}\rm{sSFR})^{-1}]$, one can see its dependence on the gas depletion time $t_{\rm dep}$ and the specific star formation rate (sSFR), both of which are expected to be redshift-dependent quantities.  For instance, some studies suggest that the typical sSFR of main sequence galaxies reaches a plateau by $z\sim 2$ \citep[e.g.,][]{Gonzalez_2010, Rodighiero_2010, Weinmann_2011}, which would result in lower gas fractions than if the sSFR steadily increases beyond $z=2$, as suggested by other studies \citep[e.g.,][]{Bouwens_2012, Stark_2013}.  Thus, for example, if in this study we underestimated the amount of gas in galaxies, this would lead to underestimating \mdust.  For a fixed stellar mass, more dust means that the quantity of UV photons per unit mass of dust would be lower, resulting in a decreased dust temperature.  It turns out that \tdust~in our model is fairly robust to changes in the gas and dust mass.  For instance, if \mdust~were higher by a factor of two for all galaxy stellar masses at all redshifts, this would decrease the values of \tdust~reported here by a factor of only $\sim 0.9$.

\subsection{Opacity, metallicity, and dust-to-gas-ratio}\label{sec:omd}
In Section \ref{sec:opacity}, we described how we use Galactic laws to model the opacity $\kappa_\nu$.  In particular, we used the \citet{Beckwith_1990} relation for long wavelengths and discussed how changes in the normalization or slope of $\kappa_\nu$ could affect the resulting dust temperatures.  We performed a series of tests to quantify these changes and found that changing $\beta$ by $\pm 1$ affects the average \tdust~by a factor of only about 1.3.  On the other hand, varying the normalization by a factor of 2 affects $\langle\tdust\rangle$  by a factor of only 1.1, on average.

While our results are robust to the opacity model, $\tdust$ is more sensitive to the prescription for galactic metallicity and the dust-to-gas ratio.  For the relationship between metallicity and DGR, we adopt the prescription of \citet{Remy-Ruyer_2014} determined from observations of galaxies at $z=0$.  For galaxies with $\mathcal{Z}>0.26\mathcal{Z}_\odot$, equation (\ref{eq:dgr}) states $\rm{DGR}/\rm{DGR}_\odot = \mathcal{Z}/\mathcal{Z}_\odot$.  This assumption is likely to break down for high-redshift, young galaxies, where the dust production sites have not yet reached equilibrium, especially if dust production by AGB stars is important \citep[e.g.,][]{Dwek_2011}.  Thus, such high-redshift galaxies would have higher DGRs than predicted by equation (\ref{eq:dgr}), which would translate into higher dust masses.

\section{Summary}\label{sec:summary}
In this paper, we have modeled the evolution of dust in galaxies, from $z=0$ to $z=9.5$, and made predictions for the dust mass and temperature as a function of galaxy stellar mass and time.  Our simple model employs empirically motivated prescriptions to determine relationships between galaxy halo mass, stellar mass, SFR, gas mass, metallicity, and dust-to-gas-ratio.

\begin{itemize}
\item Our model faithfully represents observed trends between galaxy dust and stellar mass out to $z\approx 6$.  

\item Our model predicts that the normalization between galaxy \mdust-\mstar~relation gradually decreases over time from $z=9.5$ to $z=0$, suggesting that for fixed stellar masses, galaxies in the early Universe had greater quantities of dust than modern galaxies. 
We parameterize the  \mdust-\mstar~relation as a broken power law and as a function of time.  This relationship may be useful to observers who have measurements of a galaxy's total stellar mass but are lacking observations that would provide an estimate of the dust mass.

\item In our fiducial model, in which dust, gas, and stars are homogeneously mixed together in a spherically symmetric system, the relation between galaxy dust temperature and stellar mass increases from $z=0$ to $z=6$, indicating that earlier galaxies have hotter dust.  The \tdust-\mstar~relation is not a monotonic function, but rather peaks at characteristic values of \mstar~that evolve with redshift.  The height of the peaks is sensitive to the fraction of galactic surface area covered by dust; and the exact shape of the \tdust-\mstar~relation depends on the geometry of stars and the ISM.

\item We make predictions for the observed 1.1-mm flux density, \snu, arising from dust emission in galaxies.  Our model anticipates that for a fixed galaxy stellar mass, \snu~gradually decreases with cosmic time, until $z\approx 1$, at which point it sharply rises.  There may be a population of low-mass ($\mstar\lesssim 10^9$), high-redshift ($z\gtrsim 3$) galaxies that dust as hot as, or hotter than, their more massive counterparts.  
\end{itemize} 

Given our calculations of $S_\nu$, detecting the dust emission from such low-mass, high-$z$ galaxies to determine their dust temperatures would require long integration times with current observatories, possibly making such observing programs challenging with current technology.  However, deep ALMA observations of strong lensing clusters may provide the magnification needed to measure the reemission of stellar radiation by dust in this galaxy population in a more timely fashion.  And there may be other promising ways to constrain the dust temperatures of early low-mass galaxies, for instance, by carefully modeling the contribution of their IR luminosities to the cosmic infrared background.  In massive, high-redshift galaxies, JWST has the potential to observe dust extinction of UV photons---and thus constrain dust creation and destruction---in the observed optical and infrared range of wavelengths.  Moreover, JWST observations have the potential to provide new constraints at high redshifts on the relations we use here between the SFR, metallicity, and dust-to-gas ratio.   Thus, it can be hoped that combining future JWST and ALMA observations will illuminate new aspects of the content and evolution of dust in the earliest galaxies.

\acknowledgements
Nia Imara thanks the John Harvard Distinguished Science Fellowship Program for supporting this research.  The authors are grateful to H. Jabran Zahid for valuable discussions during our work on this study.  
We also thank the referee, whose thorough reading and insightful comments helped to improve this paper.

\bibliography{dustygal}

\end{document}